\documentclass[12pt,draftcls,onecolumn]{IEEEtran}

\usepackage[T1]{fontenc}
\usepackage{graphicx}

\usepackage{amsthm,amsmath,amssymb,bm}
\usepackage{mathrsfs}
\usepackage{amsfonts}
\newtheorem{theorem}{Theorem}
\newtheorem{definition}{Definition}
\newtheorem{lemma}{Lemma}

\newtheorem{remark}{Remark}
\newtheorem{assumption}{Assumption}
\newtheorem{proposition}{Proposition}
\usepackage{subfigure}
\usepackage{algorithm}
\usepackage{algpseudocode}
\usepackage{graphics}
\usepackage{epsfig}
\usepackage{cite}
\usepackage{color}
\usepackage{booktabs}
\usepackage{verbatim}

\begin{document}
	
\title{Dynamic Clustering and Power Control for Two-Tier Wireless Federated Learning}
\author{Wei~Guo,~\IEEEmembership{Graduate Student Member,~IEEE},~Chuan~Huang,~\IEEEmembership{Member,~IEEE},~Xiaoqi~Qin,~\IEEEmembership{Member,~IEEE},~Lian~Yang,~\IEEEmembership{Member,~IEEE},~and~Wei~Zhang,~\IEEEmembership{Fellow,~IEEE}
\thanks{	
		W. Guo is with the School of Science and Engineering (SSE) and Future Network of Intelligence Institute (FNii), The Chinese University of Hong Kong, Shenzhen, China, 518172. Email: weiguo1@link.cuhk.edu.cn.
		
		C. Huang is currently with the School of Science and Engineering (SSE) and Future Network of Intelligence Institute (FNii), The Chinese University of Hong Kong, Shenzhen, China, 518172, and Peng Cheng Laboratory, Shenzhen, China, 518066. Email: huangchuan@cuhk.edu.cn.
		
		X. Qin is with the State Key Laboratory of Networking and Switching Technology, Beijing University of Posts and Telecommunications, Beijing, China, 100876. Emial: xiaoqiqin@bupt.edu.cn.
		
		L. Yang is with the Information and Communication Engineering College, University of Electronic Science and Technology of China, Chengdu, China, 611731. Email: yanglian@uestc.edu.cn.		
		
		W. Zhang is with the School of Electrical Engineering and Telecommunications, University of New South Wales, Sydney, Australia, NSW 2052. Email: w.zhang@unsw.edu.au.}}

\maketitle

\begin{abstract}
Federated learning (FL) has been recognized as a promising distributed learning paradigm to support intelligent applications at the wireless edge, where a global model is trained iteratively through the collaboration of the edge devices without sharing their data. However, due to the relatively large communication cost between the devices and parameter server (PS), direct computing based on the information from the devices may not be resource efficient. This paper studies the joint communication and learning design for the over-the-air computation (AirComp)-based two-tier wireless FL scheme, where the lead devices first collect the local gradients from their nearby subordinate devices, and then send the merged results to the PS for the second round of aggregation. We establish a convergence result for the proposed scheme and derive the upper bound on the optimality gap between the expected and optimal global loss values. Next, based on the device distance and data importance, we propose a hierarchical clustering method to build the two-tier structure. Then, with only the instantaneous channel state information (CSI), we formulate the optimality gap minimization problem and solve it by using an efficient alternating minimization method. Numerical results show that the proposed scheme outperforms the baseline ones.
\end{abstract}


\IEEEpeerreviewmaketitle
\section{Introduction}\label{sec:intro}
With the proliferation of the mobile edge devices (such as smart phones, wearable devices, and sensors), a massive amount of data has been generated at the network edge\cite{cisco}. The recent progresses in machine learning (ML) techniques \cite{KKang,YGoldberg}, fueled by the rich data, have the potential to enable a wide range of intelligence applications at the wireless edge networks, such as intelligent health \cite{JLi} and autonomous driving \cite{SRPokherel}. However, conventional ML model training methods are conducted in a centralized fashion, which requires the central server to aggregate high volume raw data obtained at the edge devices and thus inevitably causes data privacy concerns \cite{IGoodfellow}. Besides, collecting a large amount of raw data to implement the centralized learning methods in wireless scenarios inevitably consumes too much resources \cite{DChen}, and thus it is hardly practical to be employed at the wireless edge \cite{WSaad}. To overcome the aforementioned challenges, federated learning (FL) \cite{HBMcMahan,JKone,QYang} has been proposed as a promising distributed ML paradigm to protect the local data privacy and relieves the communication burden by allowing the edge devices to train the ML model locally and only upload their local gradients to the parameter server (PS) without sharing their data \cite{WYBLim,YZhao,TLi}.

Despite the above advantages of FL, deploying FL in wireless scenarios requires a large number of high-dimensional gradients and global model exchanges between the PS and the edge devices, which still suffer from the communication bottleneck, especially for the case with large number of edge devices \cite{GZhu}. To address this issue, over-the-air computation (AirComp) technique has been introduced to the wireless FL systems to support concurrently gradient uploading from the edge devices over the shared radio resources by exploiting the waveform-superposition nature of the wireless multiple access channels (MAC) \cite{BNazer,WLiu}. Despite the advantages of reducing the communication latency and enhancing the scalability of the wireless FL systems, AirComp-based gradient aggregation suffers from the distortion caused by the additive noise and channel fading of the wireless channels, which significantly degrades the FL performance \cite{XCao_1}, and comprehensive studies have been conducted to tackle this issue \cite{KYang,NZhang,GZhu_1,XCao,XFan}. In \cite{KYang}, the authors proposed a joint device selection and receiver beamforming to maximize the number of the selected devices in each iteration under the mean square error (MSE) constraint for the aggregated signals. The authors in \cite{NZhang} proposed a joint transmit power and de-noising factor optimization method to minimize the per-iteration gradient aggregation error by taking the gradient statistic into account. In \cite{GZhu_1}, the authors proposed the one-bit gradient quantization scheme at the transmitters and the AirComp majority-voting based decoding at the receiver, and analyzed the effect of noise, channel fading, and channel estimation errors on the FL performance in terms of convergence rate. In \cite{XCao}, the authors proposed an optimal transmit power allocation to minimize the expected optimality gap of the wireless FL system for both the cases with biased and unbiased gradient aggregations. In \cite{XFan}, the authors proposed a joint device selection and transmit power design to accelerate the convergence rate of the wireless FL system for both the convex and non-convex loss function cases. 

The results shown in the above researches exhibit quite similar solutions to combat the negative effects of the additive noise and channel fading, where they discard the stragglers, i.e., the devices with relatively weak channel gains, due to the limited transmit power budget. However, discarding certain amount of devices during the training process may degrade the performance of the wireless FL systems due to the insufficient training data exploitation \cite{HLiu}. Therefore, to fully exploit all the available data, one possible solution is to apply multi-tier computing techniques into the wireless FL systems, which is able to enhance the stragglers' access probability to the PS and has been reported to show better learning performance compared to the conventional FL structure with direct communications between the devices and the PS \cite{ZQu,ZLin,RJiang,MSHAbad,LLiu,RRuby}. In \cite{ZQu}, the authors studied the digital two-tier relay-assisted FL framework and proposed a partially synchronized parallel scheme to simultaneously transmit the gradients from the edge devices to relays and models from relays to the PS, which is able to reduce the training time. In \cite{ZLin}, the authors proposed a two-tier relay-assisted AirComp-based wireless FL scheme, and optimized the transmit power coefficients at the devices and relays and the de-noising factors at the PS by minimizing the MSE of the aggregated signals with alternating minimization method. Instead of introducing additional relays, the authors \cite{RJiang} proposed a cluster-based two-tier wireless FL scheme based on the method proposed in \cite{GZhu_1}, where they select a set of devices as relays in each cluster to aggregate the gradients from the devices within the same cluster and the selected devices upload the aggregated gradients based on AirComp. The results in \cite{RJiang} numerically showed that the proposed scheme achieves comparable learning performance to the ideal FL system without channel fading and additive noise. However, the aforementioned works only considered device selection for the proposed scheme without providing any convergence analysis, power control, or clustering design for the proposed scheme. In \cite{MSHAbad}, the authors investigated the gradient sparsification and periodic averaging for the two-tier wireless FL system, where devices transmit the sparsified gradients to the associated small-cell base station (SBS) for intermediate model updates and the SBSs transmit the updated models to the macro based station (MBS) after several intermediate model updates. In \cite{LLiu}, the authors investigated a two-tier client-edge-cloud wireless FL system, where the devices upload the model parameters to the associated edge servers after multiple local updates and the edge servers upload the updated model parameters to the cloud server after several model aggregations. The authors in \cite{LLiu} analyzed the convergence of the proposed two-tier FL scheme and revealed the impact of the number of the local updates and edge aggregations on the convergence. In \cite{RRuby}, the authors investigated the energy efficient resource allocation for a similar two-tier wireless FL systems as in \cite{MSHAbad,LLiu}, where the devices upload the gradients to the middle-tier aggregators and the middle-tier aggregators upload the aggregated gradients immediately to the top-tier aggregator. However, none of above works considered the joint communication and learning design, i.e., analytically characterizing the impact of communications on the learning performance, and some of them focus on only the communication aspect \cite{ZQu,ZLin,RJiang} and the others only for the learning aspect \cite{MSHAbad,LLiu,RRuby}.

This paper proposes an joint communication and learning design for an AirComp-based two-tier wireless FL scheme, which dynamically divides all the devices into different clusters across different training iterations. In particular, the subordinate devices transmit their local gradients to their associated lead device within the same cluster and all the lead devices transmit their aggregated gradients to the PS through AirComp. The main contributions of this paper are summarized as follows.
\begin{enumerate}
	\item We analyze the convergence behavior of the considered FL scheme and derive the optimality gap between the expected and optimal global loss values, which is applicable to any clustering scheme and characterize the impact of the AirComp-based two-tier gradient aggregation errors, in terms of their bias and MSE, on the optimality gap.
	\item We propose a hierarchical clustering method by adopting the minimax linkage criterion to dynamically divides the devices into a given number of clusters. In particular, we incorporate the device distance, which affects the communication aspect, and the data importance, which affects the learning performance. Then, the lead device in each cluster is selected by jointly considering its location within the corresponding cluster, its distance to the PS, and its data importance. 
	\item With the given cluster result and the instantaneous channel state information (CSI), we formulate the optimality gap minimization problem with respect to (w.r.t) the transmit power at the devices and de-noising factor at the PS. We propose an efficient algorithm based on alternating optimization to solve this problem, where the closed-form solutions for all sub optimization problems in each iteration are derived.
	
\end{enumerate}

The reminder of this paper is organized as follows. Section \ref{sec:system} presents the system model. Section \ref{sec:convergence} provides the convergence analysis results and quantifies the impact of the gradient aggregation errors on the optimality gap. Section \ref{sec:cluster_power} proposes the clustering scheme and develops an efficient alternating minimization method to solve the optimality gap minimization problem.  Numerical results are presented in Section \ref{sec:numerical}. Finally, Section \ref{sec:conclusion} concludes this paper.

{\it Notation}: $\mathbb{E}(\cdot)$ denotes the expectation operator and $\nabla$ denotes the gradient operator. The bold lower-case letter, e.g., $\mathbf{a}$, denotes a vector and the calligraphic upper-case letter, e.g., $\mathcal{A}$, denotes a set.  For a vector $\mathbf{a}$, $\|\mathbf{a}\|_2$ denotes the Euclidean norm of $\mathbf{a}$; $\mathbf{a}^{T}$ denotes the transpose of a complex vector $\mathbf{a}$. $\langle\mathbf{a},\mathbf{b}\rangle$ denotes the inner product of vectors $\mathbf{a}$ and $\mathbf{b}$. We denote a circularly symmetric complex Gaussian (CSCG) distribution with variance $\sigma^2$ by $\mathcal{CN}(0,\sigma^2)$. $\mathbb{R}$ and $\mathbb{C}$ denote the sets of real and complex numbers, respectively. For a set $\mathcal{A}$, $|\mathcal{A}|$ denotes its cardinality.

\section{System Model}\label{sec:system}
We consider a wireless FL system with one PS and a set of devices, denoted as $\mathcal{K}=\{1,\cdots, K\}$, where he PS coordinates all the devices to train a global ML model $\mathbf{w}\in\mathbb{R}^{M}$ of dimension $M$. Define the local dataset at device $k$ as $\mathcal{D}_k=\{(\mathbf{x}_{k,i},y_{k,i})\}_{i=1}^{D_k}$ with $D_k=|\mathcal{D}_k|$ data samples, where $\mathbf{x}_{k,i}\in\mathbb{R}^d$ is the $d$-dimensional input data and $y_{k,i}\in\mathbb{R}$ is the labeled output of $\mathbf{x}_{k,i}$. The goal of training the ML model is to find the optimal model parameters $\mathbf{w}^*$ such that
\begin{equation}\label{eq:FL_obj}
	\mathbf{w}^*=\arg\min_\mathbf{w} F(\mathbf{w}),
\end{equation}
where $F(\mathbf{w})$ is the global loss function defined as 
\begin{equation}\label{eq:global_loss}
	F(\mathbf{w})=\frac{1}{K}\sum_{k\in\mathcal{K}}F_k(\mathbf{w}),
\end{equation}
with $F_k(\mathbf{w})$ denoting the local loss function at device $k$ given by
\begin{equation}\label{eq:local_loss}
	F_k(\mathbf{w})=\frac{1}{D_k}\sum_{i=1}^{D_k}f(\mathbf{w};(\mathbf{x}_{k,i},y_{k,i})),
\end{equation}
with $f(\mathbf{w};(\mathbf{x}_{k,i},y_{k,i}))$ being an empirical sample-wise loss function that quantifies the loss of model $\mathbf{w}$ for data sample $(\mathbf{x}_{k,i},y_{k,i})$.
\begin{figure}[H] 
	\centering 
	\includegraphics[width=0.45\textwidth]{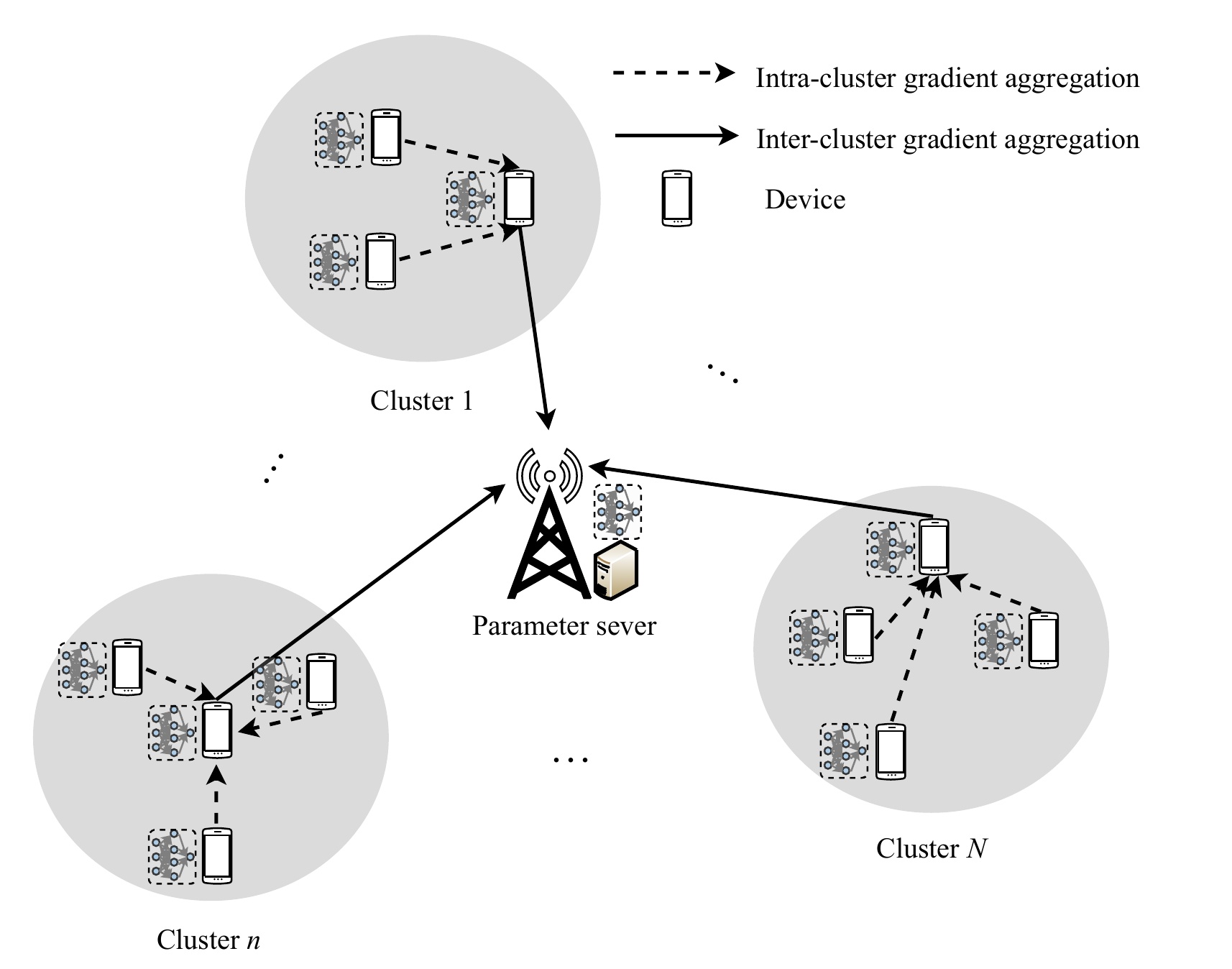} 
	\caption{AirComp-based two-tier wireless FL system with one PS, $N$ clusters, and $K$ devices.}
	\label{fig:system}
\end{figure}

This paper introduces the AirComp-based two-tier wireless FL scheme to solve problem \eqref{eq:FL_obj} in an iterative fashion. As shown in Fig. \ref{fig:system}, during the $t$-th iteration, the main idea of the proposed scheme is that we apply a clustering method to divide the all the devices into $N$ different clusters $\mathcal{C}_1^{(t)},\cdots,\mathcal{C}_N^{(t)}$, where cluster $\mathcal{C}_n^{(t)}$ contains $C_n^{(t)}=|\mathcal{C}_n^{(t)}|$ devices with $K=\sum_{n=1}^{N}C_n^{(t)}$ and $N$ is constant across different iterations, and then one lead device is selected in each cluster to perform intra-cluster gradient aggregation and the PS performs inter-cluster gradient aggregation through AirComp. Specifically, the considered scheme consecutively executes the following four steps during the $t$-th iteration:
\begin{enumerate}
	\item{\textbf{Model boradcasting:}} The PS broadcasts the global model $\mathbf{w}^{(t)}$ and all devices can successfully download the global model $\mathbf{w}^{(t)}$.
	\item{\textbf{Local Processing:}} Instead of directly computing the local gradient $\nabla F_k(\mathbf{w})$ of the local loss function with high computational cost, device $k\in\mathcal{K}$ estimates the local gradient based on randomly sampled mini-batch from its local dataset $\mathcal{D}_k$ as
	\begin{equation}\label{eq:gradient}
		\mathbf{g}_k^{(t)}=\frac{1}{m_b}\sum_{\xi_{k,i}^{(t)}\in\mathcal{B}_k^{(t)}}\nabla f\left(\mathbf{w}_k^{(t)};\left(\mathbf{x}_{k,i}^{(t)},y_{k,i}^{(t)}\right)\right),
	\end{equation}
	where $\mathcal{B}_k^{(t)}$ is the randomly sampled mini-batch with mini-batch size $|\mathcal{B}_k^{(t)}|=m_b$ and $\left(\mathbf{x}_{k,i}^{(t)},y_{k,i}^{(t)}\right)$ is the $i$-th training data in $\mathcal{B}_k^{(t)}$. 
	
	To facilitate the power control, the local gradient vectors $\mathbf{g}_k^{(t)}$ are first transformed into normalized symbols $\mathbf{s}_k^{(t)}$ with zero mean and unit variance, since the values of $\mathbf{g}_k^{(t)}$ may vary significantly \cite{GZhu,ZLin}. Let $g_{k}^{(t)}[m]$, $1\leq m\leq M$, $k\in\mathcal{K}$, denote the $m$-th element of the gradient vector $\mathbf{g}_k^{(t)}$ at device $k$, and let $s_k^{(t)}[m]$, $1\leq m\leq M$, $k\in\mathcal{K}$, denote the $m$-th element of $\mathbf{s}_k^{(t)}$. Specifically, device $k$ first computes the local gradient statistics, i.e., the mean $\bar{g}_k^{(t)}$ and variance $(\nu_k^{(t)})^2$, which are given as
	\begin{align}
		\bar{g}_k^{(t)}&=\frac{1}{M}\sum_{m=1}^{M}g_k^{(t)}[m], \label{eq:mean}\\
		(\nu_k^{(t)})^2&=\frac{1}{M}\sum_{m=1}^{M}\left(g_k^{(t)}[m]-\bar{g}_k^{(t)}[m]\right)^2,\label{eq:variance}
	\end{align}
	respectively, and uploads the quantities $\{\bar{g}_k^{(t)},\nu_k^{(t)}\}$ to the PS\footnote{We assume the uploading of the quantities $\{\bar{g}_k^{(t)},\nu_k^{(t)}\}$ to the PS in the $t$-th iteration is error-free with negligible overhead \cite{HLiu}.}. Then, the PS computes the average of the local gradient's means and variances as
	\begin{align}
		\bar{g}^{(t)} &= \frac{1}{K}\sum_{k\in\mathcal{K}}\bar{g}_k^{(t)},\label{eq:mean_avg}\\
		(\nu^{(t)})^2& =\frac{1}{K}\sum_{k\in\mathcal{K}}(\nu_k^{(t)})^2,\label{eq:variance_avg}
	\end{align}
	respectively, and broadcast the quantities $\{\bar{g}^{(t)},(\nu^{(t)})^2\}$ back to the devices for normalization. After device $k$ receives $\{\bar{g}^{(t)},(\nu^{(t)})^2\}$,  it transforms the $m$-th element of the local gradient $g_{k}^{(t)}[m]$ into the normalized symbol $s_{k}^{(t)}[m]$, i.e.,
	\begin{equation}\label{eq:normalized_symbol}
		\ s_k^{(t)}[m]\triangleq\frac{g_k^{(t)}[m]-\bar{g}^{(t)}}{\nu^{(t)}},\ 1\leq m\leq M.
	\end{equation}
	
	\item{\textbf{Intra-cluster gradient aggregations}}: In this step, the lead device aggregates the local gradients from all subordinate devices in this cluster. Let $k_n$ denote the lead device in cluster $\mathcal{C}_n^{(t)}$ and $\tilde{k}_{n}$ denote the subordinate devices (i.e., the rest devices excluding the lead device) in cluster $\mathcal{C}_n^{(t)}$. Besides, let $h_{\tilde{k}_n,k_n}^{(t)}\in\mathbb{C}$, $\tilde{k}_n, k_n\in\mathcal{C}_n^{(t)}$, denote the channel coefficient of the link between the subordinate device $\tilde{k}_n$ and its associated lead device $k_n$ in cluster $\mathcal{C}_n^{(t)}$. Throughout the paper, we consider quasi-static wireless channel model, where all the channels remain unchanged during one iteration. Then, the subordinate device $\tilde{k}_n\in\mathcal{C}_n^{(t)}$ sets its transmit sequence $\{u_{\tilde{k}_n}^{(t)}[m]:1\leq m\leq M\}$ as
	\begin{equation}\label{eq:transmit_signal}
		u_{\tilde{k}_n}^{(t)}[m] = a_{\tilde{k}_n}^{(t)}s_{\tilde{k}_n}^{(t)}[m],
	\end{equation} 
	where $a_{\tilde{k}_n}^{(t)}\in\mathbb{C}$ denotes the transmit coefficient of the subordinate device $\tilde{k}_n$. With the CSI about $h_{\tilde{k}_n,k_n}^{(t)}$ perfectly known to the subordinate device $\tilde{k}_n$,  it sets the transmit coefficient as $a_{\tilde{k}_n}^{(t)}=\sqrt{\alpha_{\tilde{k}_n}^{(t)}}e^{-j\theta_{\tilde{k}_n,k_n}^{(t)}}$ with $\theta_{\tilde{k}_n,k_n}^{(t)}$ being the phase of $h_{\tilde{k}_n,k_n}^{(t)}$ and $\alpha_{\tilde{k}_n}^{(t)}$ denoting the controllable transmit power to compensate the channel phase, which satisfies the individual transmit power constraint as
	\begin{equation}\label{eq:GCPowerBudget}
		\mathbb{E}\left[\left|u_{\tilde{k}_n}^{(t)}[m]\right|^2\right]=\alpha_{\tilde{k}_n}^{(t)}\leq P_{\tilde{k}_n}^{\rm max},
	\end{equation}
	where $P_{\tilde{k}_n}^{\rm max}$ is the maximum power budget at the subordinate device $\tilde{k}_n$. Then, the received signal in the $m$-th time slot at the lead device $k_n$ is given as
	\begin{equation}\label{eq:LDRxSignal}
		v_{k_n}^{(t)}[m]=\sum_{\tilde{k}_n\in\mathcal{C}_n^{(t)}}\bar{h}_{\tilde{k}_n,k_n}^{(t)}\sqrt{\alpha_{\tilde{k}_n}^{(t)}}s_{\tilde{k}_n}^{(t)}[m]+z_{k_n}^{(t)}[m],
	\end{equation} 
	where $\bar{h}_{\tilde{k}_n,k_n}^{(t)}$ is the magnitude of $h_{\tilde{k}_n,k_n}^{(t)}$ and $z_{k_n}^{(t)}[m]\in\mathbb{C}$ is the independent and identically distributed (i.i.d.) CSCG noise following the distribution $\mathcal{CN}(0,\sigma_{k_n}^2)$. Hence, the aggregated gradient at the lead device $k_n$ in cluster $\mathcal{C}_n^{(t)}$ becomes $\mathbf{v}_{k_n}^{(t)}=[v_{k_n}^{(t)}[1],\cdots,v_{k_n}^{(t)}[M]]^T\in\mathbb{R}^M$. In this step, we consider the case that the intra-cluster gradient aggregations for nearby clusters are performed in different time slots to avoid possible cross-cluster interference, and the corresponding latency is negligible compared to the local processing time.

	\item{\textbf{Inter-cluster gradient aggreagtion:}} After the lead device aggregates the gradients within its cluster, it transmits its aggregated gradient to the PS.  Let $b_{k_n}^{(t)}\in\mathbb{C}$, $k_n\in\mathcal{C}_n^{(t)}$, denote the transmit coefficient at the lead device $k_n$, and let $h_{k_n}^{(t)}\in\mathbb{C}$, denote the channel coefficient between the lead device $k_n$ to the PS.  Similarly, with CSI about $h_{k_n}^{(t)}$ perfectly known to the lead device $k_n$, it sets the transmit coefficient $b_{k_n}^{(t)}=\sqrt{\beta_{k_n}^{(t)}}e^{-j\theta_{k_n}^{(t)}}$ with $\theta_{k_n}^{(t)}$ being the phase of $h_{k_n}^{(t)}$ and $\beta_{k_n}^{(t)}$ denoting the controllable transmit power at the lead device $k_n$. Then, the lead device $k_n$ sets its transmit sequence  $\{u_{k_n}^{(t)}[m]:1\leq m\leq M\}$ as 
	\begin{equation}
		\begin{aligned}
			&u_{k_n}^{(t)}[m]=b_{k_n}^{(t)}v_{k_n}^{(t)}[m]\\
			=&b_{k_n}^{(t)}\left(\sum_{\tilde{k}_n\in\mathcal{C}_n^{(t)}}\bar{h}_{\tilde{k}_n,k_n}^{(t)}\sqrt{\alpha_{\tilde{k}_n}^{(t)}}s_{\tilde{k}_n}^{(t)}[m]+z_{k_n}^{(t)}[m]\right),
		\end{aligned}
	\end{equation}
	which satisfies the individual power constraint, i.e.,
	\begin{equation}\label{eq:LCPowerBudget}
		\begin{aligned}
			\mathbb{E}\left[\left|b_{k_n}^{(t)}v_{k_n}^{(t)}[m]\right|^2\right]=\beta_{k_n}^{(t)}\left(\sum_{\tilde{k}_n\in\mathcal{C}_n^{(t)}}\left(\bar{h}_{\tilde{k}_n,k_n}^{(t)}\right)^2\alpha_{\tilde{k}_n}^{(t)}+\sigma_{k_n}^2\right)\leq P_{k_n}^{\rm max}.
		\end{aligned}
	\end{equation} 
	Then, the received signal at the PS in the $(M+m)$-th time slot is given as
	\begin{equation}\label{eq:PsRxSignal}
		v^{(t)}[m]=\sum_{n=1}^{N}\sum_{k_n\in\mathcal{C}_n^{(t)}}\bar{h}_{k_n}^{(t)}\sqrt{\beta_{k_n}^{(t)}}v_{k_n}^{(t)}[m]+z^{(t)}[m],
	\end{equation}
	where $\bar{h}_{k_n}^{(t)}$ is the magnitude of $h_{k_n}^{(t)}$ and $z^{(t)}[m]\in\mathbb{C}$ is the i.i.d. CSCG noise following the distribution $\mathcal{CN}(0,\sigma^2)$. 
	
	After receiving $v^{(t)}[m]$, the PS estimates the global gradient of the global loss function by applying a de-noising factor $\zeta^{(t)}$ and a de-normalization step as \cite{GZhu} 
	\begin{align}\label{eq:estimate_model}
		\tilde{g}^{(t)}[m]=&\nu^{(t)}\left(\frac{1}{K}\zeta^{(t)}v^{(t)}[m]\right)+\bar{g}^{(t)}\nonumber\\
		=&\frac{1}{K}\sum_{n=1}^{N}\sum_{\tilde{k}_n\in\mathcal{C}_n^{(t)}}\nu^{(t)}\zeta^{(t)}\bar{h}_{k_n}^{(t)}\sqrt{\beta_{k_n}^{(t)}}\bar{h}_{\tilde{k}_n,k_n}^{(t)}\sqrt{\alpha_{\tilde{k}_n}^{(t)}}s_{\tilde{k}_n}^{(t)}[m]\nonumber\\
		&+\frac{1}{K}\sum_{n=1}^{N}\sum_{k_n\in\mathcal{C}_n^{(t)}}\zeta^{(t)}\nu^{(t)}\bar{h}_{k_n}^{(t)}\sqrt{\beta_{k_n}^{(t)}}z_{k_n}^{(t)}[m]+\frac{1}{K}\zeta^{(t)}\nu^{(t)}z^{(t)}[m]+\bar{g}^{(t)}\nonumber\\
		=&\frac{1}{K}\sum_{n=1}^{N}\sum_{\tilde{k}_n\in\mathcal{C}_n^{(t)}}\zeta^{(t)}\bar{h}_{k_n}^{(t)}\sqrt{\beta_{k_n}^{(t)}}\bar{h}_{\tilde{k}_n,k_n}^{(t)}\sqrt{\alpha_{\tilde{k}_n}^{(t)}}\left(g_{\tilde{k}_n}^{(t)}[m]-\bar{g}^{(t)}[m]\right)\nonumber\\
		&+\frac{1}{K}\sum_{n=1}^{N}\sum_{k_n\in\mathcal{C}_n^{(t)}}\zeta^{(t)}\nu^{(t)}\bar{h}_{k_n}^{(t)}\sqrt{\beta_{k_n}^{(t)}}z_{k_n}^{(t)}[m]+\frac{1}{K}\zeta^{(t)}\nu^{(t)}z^{(t)}[m]+\bar{g}^{(t)},
	\end{align}
	where $\bar{g}^{(t)}$ and $\nu^{(t)}$ is the de-normalization term that is related to the local gradient means and variances and defined in \eqref{eq:mean_avg} and \eqref{eq:variance_avg}, respectively . 
	
	Finally, after collecting $\tilde{\mathbf{g}}^{(t)}=[\tilde{g}^{(t)}[1],\cdots,\tilde{g}^{(t)}[M]]$, the PS updates the global model as 
	\begin{equation}\label{eq:global_model}
		\mathbf{w}^{(t+1)}=\mathbf{w}^{(t)}-\gamma\tilde{\mathbf{g}}^{(t)},
	\end{equation}
	where $\gamma$ is the learning rate. Let $\bar{\mathbf{g}}^{(t)}=\frac{1}{K}\sum_{k=1}^{K}\mathbf{g}_k^{(t)}$ denote the desired aggregated gradient from \eqref{eq:global_loss}. In this case, the total communication error caused by the AirComp-based two-tier gradient aggregation is given as
	\begin{equation}\label{eq:error}
		\begin{aligned} 
			\bm{\varepsilon}^{(t)}=\tilde{\mathbf{g}}^{(t)}-\bar{\mathbf{g}}^{(t)}=\tilde{\mathbf{g}}^{(t)}-\frac{1}{K}\sum_{k=1}^{K}\mathbf{g}_k^{(t)},
		\end{aligned}
	\end{equation}
	where the $m$-th entry of $\bm{\varepsilon}^{(t)}$ is given as
	\begin{equation}\label{eq:error_element}
		\begin{aligned}
			\varepsilon^{(t)}[m]=&\frac{1}{K}\sum_{n=1}^{N}\sum_{\tilde{k}_n\in\mathcal{C}_n^{(t)}}\left(\zeta^{(t)}\bar{h}_{k_n}^{(t)}\sqrt{\beta_{k_n}^{(t)}}\bar{h}_{\tilde{k}_n,k_n}^{(t)}\sqrt{\alpha_{\tilde{k}_n}^{(t)}}-1\right)\left(g_{\tilde{k}_n}^{(t)}[m]-\bar{g}_{\tilde{k}_n}^{(t)}\right)\\
			&-\frac{1}{K}\sum_{n=1}^{N}\sum_{k_n\in\mathcal{C}_n^{(t)}}\left(g_{k_n}^{(t)}[m]-\bar{g}_{k_n}^{(t)}\right)+\frac{1}{K}\sum_{n=1}^{N}\sum_{k_n\in\mathcal{C}_n^{(t)}}\zeta^{(t)}\nu^{(t)}\bar{h}_{k_n}^{(t)}\sqrt{\beta_{k_n}^{(t)}}z_{k_n}^{(t)}[m]\\
			&+\frac{1}{K}\zeta^{(t)}\nu^{(t)}z^{(t)}[m].
		\end{aligned}
	\end{equation}
	Denote $\Delta\mathbf{g}_k^{(t)}=[g_k^{(t)}[1]-\bar{g}^{(t)},\cdots,g_k^{(t)}[M]-\bar{g}^{(t)}]^T$, $\mathbf{z}_{k_n}^{(t)}=[z_{k_n}^{(t)}[1],\cdots,z_{k_n}^{(t)}[M]]^T$, and $\mathbf{z}^{(t)}=[z^{(t)}[1],\cdots,z^{(t)}[M]]^T$. Then, \eqref{eq:error} is equivalent to
	\begin{align}\label{eq:error_new}
		\bm{\varepsilon}^{(t)}=&\frac{1}{K}\sum_{n=1}^{N}\sum_{\tilde{k}_n\in\mathcal{C}_n^{(t)}}\left(\zeta^{(t)}\bar{h}_{k_n}^{(t)}\sqrt{\beta_{k_n}^{(t)}}\bar{h}_{\tilde{k}_n,k_n}^{(t)}\sqrt{\alpha_{\tilde{k}_n}^{(t)}}-1\right)\Delta\mathbf{g}_{\tilde{k}_n}^{(t)}\nonumber\\
		&+\frac{1}{K}\sum_{n=1}^{N}\sum_{k_n\in\mathcal{C}_n^{(t)}}\zeta^{(t)}\nu^{(t)}\bar{h}_{k_n}^{(t)}\sqrt{\beta_{k_n}^{(t)}}\mathbf{z}_{k_n}^{(t)}+\frac{1}{K}\zeta^{(t)}\nu^{(t)}\mathbf{z}^{(t)}-\frac{1}{K}\sum_{n=1}^{N}\sum_{k_n\in\mathcal{C}_n^{(t)}}\Delta\mathbf{g}_{k_n}^{(t)}.
	\end{align}
	Therefore, \eqref{eq:global_model} can be rewritten as 
	\begin{equation}\label{eq:new_glob_model}
		\mathbf{w}^{(t+1)}=\mathbf{w}^{(t)}-\gamma(\bar{\mathbf{g}}^{(t)}+\bm{\varepsilon}^{(t)}).
	\end{equation}
\end{enumerate}

\begin{remark}
	Unlike many existing literatures \cite{XCao_1,XFan}, which do not consider the gradient normalization or directly measure the learning performance by aggregation error w.r.t. the normalized symbol $\mathbf{s}_k^{(t)}$ \cite{KYang,NZhang,ZLin}, we take both the normalization and de-normalization steps \eqref{eq:normalized_symbol} and \eqref{eq:estimate_model} into account, and explicitly compute the aggregation error $\bm{\varepsilon}^{(t)}$ \eqref{eq:error_new} w.r.t. the gradients $\mathbf{g}_k^{(t)}$. Hence, with the derived aggregation error $\bm{\varepsilon}^{(t)}$, we are able to analyze the impact of the AirComp-based  two-tier gradient aggregations on the convergence behavior of the wireless FL algorithm as detailed in Section \ref{subsec:result_convergence}.
\end{remark}

\section{Convergence Analysis}\label{sec:convergence}
In this section, we analyze the convergence behavior of the considered AirComp-based two-tier wireless FL algorithm presented in Section \ref{sec:system} for the general smooth non-convex learning problems. We first present the preliminaries, and then present the theoretical results on convergence for the AirComp-based two-tier wireless FL algorithm in terms of the optimality gap between the expected and optimal global loss values.

\subsection{Preliminaries}\label{subsec:pre_convergence}
First, we make the following standard assumptions that are commonly adopted in the convergence analysis in the literature \cite{XLi,GZhu,GZhu_1,XCao}.
\begin{assumption}[\textbf{Smoothness}]\label{as:smooth}
	The global loss function defined in \eqref{eq:global_loss} is differentiable and the gradient is uniformly Lipschitz continuous with a positive Lipschitz constant $L$, i.e., $\forall \mathbf{v},\mathbf{w}\in\mathbb{R}^d$,
	\begin{equation}\label{eq:smooth}
		\|\nabla F(\mathbf{v})-\nabla F(\mathbf{w})\|_2\leq L\|\mathbf{v}-\mathbf{w}\|_2,
	\end{equation}
	which is equivalent to 
	\begin{equation}\label{eq:smooth_1}
		F(\mathbf{v})\leq F(\mathbf{w})+(\mathbf{v}-\mathbf{w})^T\nabla F(\mathbf{w})+\frac{L}{2}\|\mathbf{v}-\mathbf{w}\|_2^2.
	\end{equation}
\end{assumption}
\begin{assumption}[\textbf{Bounded Variance}]\label{as:bounded_sgd}
	The obtained local gradient $\mathbf{g}_k$ is an independent and unbiased estimation for the full-batch gradient $\nabla F(\mathbf{w})$ with element-wise bounded variance, i.e.,
	\begin{align}
		&\mathbb{E}[\mathbf{g}_k]=\nabla F(\mathbf{w}),\ \forall k\in\mathcal{K},\label{eq:unbiasedSGD}\\
		&\mathbb{E}[(g_k[m]-\nabla F(w[m])^2]\leq\frac{\delta_m^2}{m_b},\ \forall k\in\mathcal{K},1\leq m\leq M,\label{eq:boundedSGD}
	\end{align}
	where $g_k[m]$ and $\nabla F(w[m])$ are the $m$-th element of $\mathbf{g}_k$ and $\nabla F(\mathbf{w})$, respectively, $\bm{\delta}=[\delta_1,\cdots,\delta_M]$ is a non-negative constant vector. 
\end{assumption}

\subsection{Convergence Results}\label{subsec:result_convergence}
Based on Assumptions \ref{as:smooth} and \ref{as:bounded_sgd} and the recasted global model \eqref{eq:new_glob_model}, we are ready to present the convergence result in the following theorem.
\begin{theorem}\label{th:opt_gap}
	Consider the scenarios when Assumptions \ref{as:smooth} and \ref{as:bounded_sgd} are valid and set the learning rate $\gamma<\frac{1}{2L}$, the optimality gap between the expected and optimal global loss function values is upper bounded by 
	\begin{equation}\label{eq:th2}
		\begin{aligned}
			&\mathbb{E}\left[F(\mathbf{w}^{(T+1)})\right]-F^*\\
			\leq& \eta^T\mathbb{E}[F(\mathbf{w}^{(1)})-F^*]+\sum_{t=1}^T\eta^{T-t}\frac{L\gamma^2}{m_b}\|\bm{\delta}\|_2^2+\sum_{t=1}^T\eta^{T-t}\left(\frac{\gamma}{2}\left\|\mathbb{E}\left[\bm{\varepsilon}^{(t)}\right]\right\|_2^2+L\gamma^2\mathbb{E}\left[\|\bm{\varepsilon}^{(t)}\|_2^2\right]\right)
		\end{aligned}
	\end{equation}
	where $\eta=2L^2\gamma^2-L\gamma+1$, $T$ is the total training iterations, $\mathbf{w}^{(1)}$ is the initial global model, and $F^*$ is the optimal loss function value. 
\end{theorem}
\begin{IEEEproof}
	Please see Appendix. \ref{appen:th1_proof}.
\end{IEEEproof}
\begin{remark}\label{rem:th_insight}
	The optimality gap derived in the right hand side (RHS) of \eqref{eq:th2} consists of four terms related to the initial gap $\mathbb{E}[F(\mathbf{w}^{(1)})-F^*]$, the gradient variance $\frac{L\gamma^2}{m_b}\|\bm{\delta}\|_2^2$, and the AirComp-based two-tier gradient aggregation error $\frac{\gamma}{2}\left\|\mathbb{E}\left[\bm{\varepsilon}^{(t)}\right]\right\|_2^2+L\gamma^2\mathbb{E}\left[\|\bm{\varepsilon}^{(t)}\|_2^2\right]$. Given the learning rate $\gamma<\frac{1}{2L}$, we have $0<\eta=2L^2\gamma^2-L\gamma+1<1$. Hence, as $T\rightarrow\infty$, the first term $\eta^T\mathbb{E}[F(\mathbf{w}^{(1)})-F^*]$ diminishes to zero. If we further choose the learning rate sufficiently small, the rest terms in the RHS of \eqref{eq:th2} can also diminish to zero. In this case, the wireless FL converges exactly to the optimal point. However, in practice, neither $T$ can go to infinity nor the learning rate is always small enough, which makes the wireless FL end up to a point in the  neighborhood of the optimal point when terminated. Besides, $\mathbb{E}\left[\bm{\varepsilon}^{(t)}\right]$ and $\mathbb{E}\left[\|\bm{\varepsilon}^{(t)}\|_2^2\right]$ in \eqref{eq:th2} are the bias and MSE of the globally aggregated gradient at the $t$-th iteration, respectively, where the expectations are taken w.r.t. the channel noise. Obviously, the squared bias $\|\mathbb{E}\left[\bm{\varepsilon}^{(t)}\right]\|_2^2$ and MSE $\mathbb{E}\left[\|\bm{\varepsilon}^{(t)}\|_2^2\right]$ are both determined by the transmit power values $\beta_{k_n}^{(t)}$ and $\alpha_{\tilde{k}_n}^{(t)}$ and de-noising factor $\zeta^{(t)}$ with given cluster results $\{\mathcal{C}_1^{(t)},\cdots,\mathcal{C}_N^{(t)}\}$ at the $t$-th iteration.
\end{remark}
\begin{remark}\label{rem:generality}
	The convergence result obtained in Theorem \eqref{th:opt_gap} is a quite general result for the considered two-tier wireless FL scheme satisfying Assumptions \ref{as:smooth}-\ref{as:bounded_sgd}, as long as we can compute the aggregation error w.r.t. the gradients $\mathbf{g}_k^{(t)}$ as \eqref{eq:new_glob_model}. Hence, Theorem \ref{th:opt_gap} is applicable to any clustering result and power and de-noising factor allocation across in the whole training process.
\end{remark}

\section{Clustering and Optimality Gap Minimization}\label{sec:cluster_power}
In this section, we first introduce the distance and data importance aware dynamic cluster method to all divide the devices into several clusters at different iterations. Then, we present the joint transmit power and de-nosing factor control to minimize the derived optimality gap \eqref{eq:th2}. 

\subsection{Distance and Data Importance Aware Clustering}\label{subsec:cluster}
To cluster the devices, we adopt the hierarchical clustering method with the minimax linkage criterion \cite{JBien,MYemini}. However, conventional clustering methods only consider the distances between the devices to determine how to group all the devices into multiple clusters. In our proposed scheme, we further incorporate the data importance to the clustering objectives such that both the communication and the learning performance can be considered for the considered scheme. Let $I_k^{(t)}$ denote the data importance for the $k$-th device at the $t$-th iteration, which is measured in different ways for various learning tasks \cite{YHe,JRen}. In this paper, we focus on the learning task of training convolutional neural network (CNN) models for image classification. In this case, a suitable measure for the date importance is entropy, which is computed as \cite{DLiu,DLiu2, AHolub} 
\begin{equation}\label{eq:entropy}
	I_k^{(t)}=-\frac{1}{D_k}\sum_{i\in\mathcal{D}_k}\sum_{\bar{y}_{k,i}\in\mathcal{Y}}\mathbb{P}_{\mathbf{w}^{(t)}}(\bar{y}_{k,i}|\mathbf{x}_{k,i})\log \mathbb{P}_{\mathbf{w}^{(t)}}(\bar{y}_{k,i}|\mathbf{x}_{k,i}),
\end{equation} 
where $\bar{y}_{k,i}$ is the prediction label for data input $\mathbf{x}_{k,i}$, $\mathcal{Y}$ is the set of all labels, and $\mathbb{P}_{\mathbf{w}^{(t)}}(\bar{y}_{k}|\mathbf{x}_{k,i})$ is the prediction probability of input $\mathbf{x}_{k,i}$ and output $\bar{y}_{k,i}$ with given model parameter $\mathbf{w}^{(t)}$. Note that $I_k^{(t)}$ in \eqref{eq:entropy} is non-negative, and the larger $I_k^{(t)}$ is, the more important for the data at the device $k$ to update the model $\mathbf{w}^{(t)}$.

Let $\mathtt{dist}(\cdot,\cdot):\mathbb{R}^2\times\mathbb{R}^2\rightarrow\mathbb{R}$ be the Euclidean distance between any two devices, and let $\mathcal{P}$ be a set of points in $\mathbb{R}^2$, i.e., the locations of the devices in $\mathcal{K}$. We then define the following concepts \cite{MYemini}:
\begin{definition}{(Distance of a point in a set)}
	The distance of point $p_i\in\mathcal{P}$ in the set $\mathcal{P}$  is defined as $r(p_i,\mathcal{P})=\max_{p_j\in\mathcal{P}}\mathtt{dist}(p_i,p_j)$.
\end{definition}
\begin{definition}{(Minimax distance of a set)}
	The minimax distance of the set $\mathcal{P}$ is defined as $r(\mathcal{P})=\min_{p_i\in\mathcal{P}}r(p_i,\mathcal{P})$.
\end{definition}
\begin{definition}{(Joint distance and data importance objective)}
	The joint distance and data importance objective of of a set $\mathcal{P}$ is defined as $\mathtt{obj}(\mathcal{P})=r(\mathcal{P})+\varrho\max_{p_i\in\mathcal{P}}I_{p_i}^{(t)}$, where $\varrho$ is a coefficient to balance the distance and data importance.
\end{definition}
\begin{definition}{(Minimax linkage)}\label{def:linkage}
	The minimax linkage between two sets of points $\mathcal{P}_1$ and $\mathcal{P}_2$ in $\mathbb{R}^2$ is defined as $\mathtt{link}(\mathcal{P}_1,\mathcal{P}_2)=\mathtt{obj}(\mathcal{P}_1\cup\mathcal{P}_2)$.
\end{definition}

Following the similar procedure to Algorithm 1 in \cite{MYemini} by utilizing the proposed the minimax linkage in Definition \ref{def:linkage}, we can efficiently divide the devices into $N$ clusters. After the clusters being formed, we need to further determine the lead device for each cluster. Specifically, in cluster $\mathcal{C}_n^{(t)}$, let $D_{i_n}^{(t)}$ denote the average distance between the device $i_n$ and all the rest devices within the same cluster, i.e., 
\begin{equation}\label{eq:avg_dist}
	D_{i_n}^{(t)}=\frac{1}{C_n^{(t)}-1}\sum_{j_n\in\mathcal{C}_n^{(t)},j_n\neq i_n}\mathtt{dist}(p_{j_n},p_{i_n}),\ i_n\in\mathcal{C}_n^{(t)},
\end{equation}
and let $\hat{D}_{i_n}^{(t)}$ denote the distance between device $i_n$ to the PS. Then, the lead client $k_n$ in cluster $\mathcal{C}_n^{(t)}$ is chosen according to the following rule:
\begin{equation}\label{eq:lead_selection}
	k_n =\arg\min_{i_n\in\mathcal{C}_n^{(t)}}D_{i_n}^{(t)}+\varrho_1\hat{D}_{i_n}^{(t)}+\varrho_2I_{i_n}^{(t)},
\end{equation}
where $\varrho_1$ and $\varrho_2$ are the coefficients to balance the device's location in the cluster, distance to the PS, and data importance.

\subsection{Optimality Gap Minimization}\label{subsec:optimization}
In this subsection, we consider the transmit power and de-noising factor control to minimize the optimality gap derived in Theorem \ref{th:opt_gap} to enhance the learning performance of the considered scheme. By discarding the constant terms in the RHS of \eqref{eq:th2}, i.e., the first two terms related to the initial gap $\mathbb{E}[F(\mathbf{w}^{(1)})-F^*]$ and the gradient variance $\frac{L\gamma^2}{m_b}\|\bm{\delta}\|_2^2$, we only need to minimize
\begin{equation}\label{eq:obj}
	\sum_{t=1}^{T}\left(A_t\left\|\mathbb{E}\left[\bm{\varepsilon}^{(t)}\right]\right\|_2^2+B_t\mathbb{E}\left[\|\bm{\varepsilon}^{(t)}\|_2^2\right]\right),
\end{equation}
where $A_t=\frac{\eta^{T-r}\gamma}{2}, B_t=L\gamma^2\eta^{T-t},1\leq t\leq T$.  Based on the definition of $\bm{\varepsilon}^{(t)}$ in \eqref{eq:error_new}, we further bound the squared bias and MSE of the globally aggregated gradient as
\begin{equation}\label{eq:bias_bound}
	\begin{aligned}
		\|\mathbb{E}[\bm{\varepsilon}^{(t)}]\|_2^2=&\left\|\mathbb{E}\left[\frac{1}{K}\sum_{n=1}^{N}\sum_{\tilde{k}_n\in\mathcal{C}_n^{(t)}}\left(\zeta^{(t)}\bar{h}_{k_n}^{(t)}\sqrt{\beta_{k_n}^{(t)}}\bar{h}_{\tilde{k}_n,k_n}^{(t)}\sqrt{\alpha_{\tilde{k}_n}^{(t)}}-1\right)\Delta\mathbf{g}_{\tilde{k}_n}^{(t)}-\frac{1}{K}\sum_{n=1}^{N}\sum_{k_n\in\mathcal{C}_n^{(t)}}\Delta\mathbf{g}_{k_n}^{(t)}\right]\right\|_2^2\\
		\leq&\frac{\sum_{k\in\mathcal{K}}\|\Delta\mathbf{g}_k^{(t)}\|_2^2}{K^2}\sum_{n=1}^{N}\sum_{\tilde{k}_n\in\mathcal{C}_n^{(t)}}\left(\zeta_2^{(t)}\bar{h}_{k_n}^{(t)}\sqrt{\beta_{k_n,2}^{(t)}}\bar{h}_{\tilde{k}_n,k_n}^{(t)}\sqrt{\alpha_{\tilde{k}_n}^{(t)}}-1\right)^2+\frac{N\sum_{k\in\mathcal{K}}\|\Delta\mathbf{g}_k^{(t)}\|_2^2}{K^2},
	\end{aligned}
\end{equation}
\begin{equation}\label{eq:mse_bound}
	\begin{aligned}
		\mathbb{E}\left[\|\bm{\varepsilon}^{(t)}\|_2^2\right]=&\mathbb{E}\left[\left\|\frac{1}{K}\sum_{n=1}^{N}\sum_{\tilde{k}_n\in\mathcal{C}_n^{(t)}}\left(\zeta^{(t)}\bar{h}_{k_n}^{(t)}\sqrt{\beta_{k_n}^{(t)}}\bar{h}_{\tilde{k}_n,k_n}^{(t)}\sqrt{\alpha_{\tilde{k}_n}^{(t)}}-1\right)\Delta\mathbf{g}_{\tilde{k}_n}^{(t)}-\frac{1}{K}\sum_{n=1}^{N}\sum_{k_n\in\mathcal{C}_n^{(t)}}\Delta\mathbf{g}_{k_n}^{(t)}\right\|_2^2\right]\\
		&+\frac{M\sum_{n=1}^{N}\sum_{k_n\in\mathcal{C}_n^{(t)}}(\zeta^{(t)})^2(\nu^{(t)})^2\left(\bar{h}_{k_n}^{(t)}\right)^2\beta_{k_n}^{(t)}\sigma_{k_n}^2+M(\nu^{(t)})^2\zeta^{(t)})^2\sigma^2}{K^2}\\
		\leq&\frac{\sum_{k\in\mathcal{K}}\|\Delta\mathbf{g}_k^{(t)}\|_2^2}{K^2}\sum_{n=1}^{N}\sum_{\tilde{k}_n\in\mathcal{C}_n^{(t)}}\left(\zeta^{(t)}\bar{h}_{k_n}^{(t)}\sqrt{\beta_{k_n}^{(t)}}\bar{h}_{\tilde{k}_n,k_n}^{(t)}\sqrt{\alpha_{\tilde{k}_n}^{(t)}}-1\right)^2+\frac{N\sum_{k\in\mathcal{K}}\|\Delta\mathbf{g}_k^{(t)}\|_2^2}{K^2}\\
		&+\frac{M\sum_{n=1}^{N}\sum_{k_n\in\mathcal{C}_n^{(t)}}(\zeta^{(t)})^2(\nu^{(t)})^2\left(\bar{h}_{k_n}^{(t)}\right)^2\beta_{k_n}^{(t)}\sigma_{k_n}^2+M(\nu^{(t)})^2\zeta^{(t)})^2\sigma^2}{K^2},
	\end{aligned}
\end{equation}
where the inequalities in \eqref{eq:bias_bound} and \eqref{eq:mse_bound} follow the Cauchy's inequality. 

Then, with the instantaneous CSI and given the dynamic cluster results in each iteration, we can minimize $A_t\left\|\mathbb{E}\left[\bm{\varepsilon}^{(t)}\right]\right\|_2^2+B_t\mathbb{E}\left[\|\bm{\varepsilon}^{(t)}\|_2^2\right]$ in \eqref{eq:obj} iteration by iteration under the individual transmit power constraints. By substituting \eqref{eq:bias_bound} and \eqref{eq:mse_bound} into the aforementioned expression and dropping the iteration index $t$ for simplicity, we obtain the objective function as
\begin{align}\label{eq:new_obj}
		&f\left(\{\alpha_{\tilde{k}_n}\},\{\beta_{k_n}\},\zeta\right)=A\left\|\mathbb{E}\left[\bm{\varepsilon}\right]\right\|_2^2+B\mathbb{E}\left[\|\bm{\varepsilon}\|_2^2\right]\nonumber\\
		=&\frac{(A+B)\sum_{k\in\mathcal{K}}\|\Delta\mathbf{g}_k\|_2^2}{K^2}\sum_{n=1}^{N}\sum_{\tilde{k}_n\in\mathcal{C}_n}\left(\zeta\bar{h}_{k_n}\sqrt{\beta_{k_n}}\bar{h}_{\tilde{k}_n,k_n}\sqrt{\alpha_{\tilde{k}_n}}-1\right)^2\nonumber\\
		&+\frac{MB\left(\sum_{n=1}^{N}\sum_{k_n\in\mathcal{C}_n}\zeta^2\nu^2\bar{h}_{k_n}^2\beta_{k_n}\sigma_{k_n}^2+\nu^2\zeta^2\sigma^2\right)}{K^2}\nonumber\\
		\triangleq&c_1\sum_{n=1}^{N}\sum_{\tilde{k}_n\in\mathcal{C}_n}\left(\zeta\bar{h}_{k_n}\sqrt{\beta_{k_n}}\bar{h}_{\tilde{k}_n,k_n}\sqrt{\alpha_{\tilde{k}_n}}-1\right)^2+c_2\left(\sum_{n=1}^{N}\sum_{k_n\in\mathcal{C}_n}\zeta^2\nu^2\bar{h}_{k_n}^2\beta_{k_n}\sigma_{k_n}^2+\zeta^2\nu^2\sigma^2\right),
\end{align}
where $c_1=\frac{(A+B)\sum_{k\in\mathcal{K}}\|\Delta\mathbf{g}_k\|_2^2}{K^2}$ and $c_2=\frac{MB}{K^2}$. Hence, the recasted optimality gap minimization problem under the individual transmit power constraints \eqref{eq:GCPowerBudget} and \eqref{eq:LCPowerBudget} is formulated as
\begin{subequations}\label{opt:problem}
	\begin{align} 			   
			\underset{\{\{\alpha_{\tilde{k}_n}\},\{\beta_{k_n}\},\zeta\}}{\text{min}}\ \ &f\left(\{\alpha_{\tilde{k}_n}\},\{\beta_{k_n}\},\zeta\right)\label{opt:obj}\\
			{\text{s. t.}}\ \ &\alpha_{\tilde{k}_n}\leq P_{\tilde{k}_n}^{\rm max},\ \tilde{k}_n\in\mathcal{K}\setminus\mathcal{K}_N,\label{opt:ak_constraint}\\
			&\beta_{k_n}\left(\sum_{\tilde{k}_n\in\mathcal{C}_n}\bar{h}_{\tilde{k}_n,k_n}^2\alpha_{\tilde{k}_n}+\sigma_{k_n}^2\right)\leq P_{k_n}^{\rm max},\nonumber\\
			&k_n\in\mathcal{C}_n,\forall n,\label{opt:bn_constraint}\\
			&\zeta>0.\label{opt:zeta_constraint}
		\end{align}
\end{subequations} 

Obviously, problem \eqref{opt:problem} is non-convex since the transmit power values $\{\alpha_{\tilde{k}_n}\}$ and $\{\beta_{k_n}\}$, and the de-noising factors $\zeta$ are coupled together in the objective function \eqref{opt:obj}. Hence, we propose an alternating minimization method to alternately optimize the transmit power values and de-noising factor. 
\begin{itemize}
	\item{\textbf{Optimizing $\{\alpha_{\tilde{k}_n}\}$ for given $\{\{\beta_{k_n}\},\zeta\}$:}} Then, problem \eqref{opt:problem} is simplified as
	\begin{subequations}\label{opt_alpha:problem}
		\begin{align} 			   
			\underset{\{\alpha_{\tilde{k}_n}\}}{\text{min}}\ \ &c_1\sum_{n=1}^{N}\sum_{\tilde{k}_n\in\mathcal{C}_n}\left(\zeta\bar{h}_{k_n}\sqrt{\beta_{k_n}}\bar{h}_{\tilde{k}_n,k_n}\sqrt{\alpha_{\tilde{k}_n}}-1\right)^2\label{opt_alpha:obj}\\
			{\text{s. t.}}\ \ &\alpha_{\tilde{k}_n}\leq P_{\tilde{k}_n}^{\rm max},\ \tilde{k}_n\in\mathcal{C}_n,\forall n,\label{opt_alpha:ak_constraint}\\
			&\sum_{\tilde{k}_n\in\mathcal{C}_n}\bar{h}_{\tilde{k}_n,k_n}^2\alpha_{\tilde{k}_n}+\sigma_{k_n}^2\leq\frac{P_{k_n}^{\rm max}}{\beta_{k_n}},\forall n. \label{opt_alpha:bn_constraint}
		\end{align}
	\end{subequations} 
	By applying the change of the variables $\bar{\alpha}_{\tilde{k}_n}\triangleq\sqrt{\alpha_{\tilde{k}_n}}$, problem \eqref{opt_alpha:problem} is equivalent to
	\begin{subequations}\label{opt_alpha1:problem}
		\begin{align} 			   
			\underset{\{\bar{\alpha}_{\tilde{k}_n}\}}{\text{min}}\ \ &c_1\sum_{n=1}^{N}\sum_{\tilde{k}_n\in\mathcal{C}_n}\left(\zeta\bar{h}_{k_n}\sqrt{\beta_{k_n}}\bar{h}_{\tilde{k}_n,k_n}\bar{\alpha}_{\tilde{k}_n}-1\right)^2\label{opt_alpha1:obj}\\
			{\text{s. t.}}\ \ &\bar{\alpha}_{\tilde{k}_n}\leq\sqrt{P_{\tilde{k}_n}^{\rm max}},\ \tilde{k}_n\in\mathcal{C},\forall n,\label{opt_alpha1:ak_constraint}\\
			&\sum_{\tilde{k}_n\in\mathcal{C}_n}\bar{h}_{\tilde{k}_n,k_n}^2\bar{\alpha}_{\tilde{k}_n}^2+\sigma_{k_n}^2\leq \bar{P}_{k_n}^{\rm max},\forall n,\label{opt_alpha1:bn_constraint}
		\end{align}
	\end{subequations} 
	with $\bar{P}_{k_n}^{\rm max}=P_{k_n}^{\rm max}/\beta_{k_n}$. Obviously, problem \eqref{opt_alpha1:problem} is a convex problem, and thus we obtain the optimal solution to problem \eqref{opt_alpha:problem} in the following proposition.
	\begin{proposition}\label{prop:opt_alpha}
		The optimal solution $\alpha_{\tilde{k}_n}^{\rm opt}$ to problem \eqref{opt_alpha:problem} is given as
		\begin{equation}\label{eq:optimal_alpha}
			\alpha_{\tilde{k}_n}^{\rm opt}=\min\left[\left(\frac{c_1\zeta\bar{h}_{k_n}\sqrt{\beta_{k_n}}\bar{h}_{\tilde{k}_n,k_n}}{c_1\zeta^2\bar{h}_{k_n}^2\beta_{k_n}\bar{h}_{\tilde{k}_n,k_n}^2+\mu_{n}^{\rm opt}\bar{h}_{\tilde{k}_n,k_n}^2}\right)^2, P_{\tilde{k}_n}^{\rm max}\right],\ \tilde{k}_n\in\mathcal{C}_n,\forall n,
		\end{equation}
		where $\mu_{n}^{\rm opt}$'s are non-negative and satisfies the following complementary slackness conditions:
		\begin{equation}\label{eq:optimal_mu}
			\mu_{n}^{\rm opt}\left(\sum_{\tilde{k}_n\in\mathcal{C}_n}\bar{h}_{\tilde{k}_n,k_n}^2\alpha_{\tilde{k}_n}^{\rm opt}+\sigma_{k_n}^2-\bar{P}_{k_n}^{\rm max}\right)=0,\forall n.
		\end{equation}
	\end{proposition}
	\begin{IEEEproof}
		Please see Appendix. \ref{appen:prop_alpha}.
	\end{IEEEproof}
	\begin{remark}\label{rem:channel_inversion}
		Proposition \ref{prop:opt_alpha} indicates that the optimal transmit power $\alpha_{\tilde{k}_n}^{\rm opt}$ at the subordinate devices $\tilde{k}_n\in\mathcal{C}_n$, $\forall n$, exhibits a regularized channel inversion structure with the regularization component $\mu_{n}^{\rm opt}\bar{h}_{\tilde{k}_n,k_n}^2$ related to the maximum power constraint at the lead device $k_n$. For the special case that the power budgets at the devices are sufficiently large, the optimal dual variable satisfies $\mu_{n}^{\rm opt}=0,\forall n$ and the optimal transmit power $\alpha_{\tilde{k}_n}^{\rm opt}$ is asymptotically given as $\alpha_{\tilde{k}_n}^{\rm opt}=\frac{1}{\zeta\bar{h}_{k_n}\sqrt{\beta_{k_n}}\bar{h}_{\tilde{k}_n,k_n}}$, which is the composite channel inversion power control scheme.
	\end{remark}

	\item {\textbf{Optimizing $\{\beta_{k_n}\}$ for given $\{\{\alpha_{\tilde{k}_n}\},\zeta\}$:}} Then, problem \eqref{opt:problem} is simplified as
	\begin{subequations}\label{opt_2ndbeta:problem}
		\begin{align} 			   
			\underset{\{\beta_{k_n}\}}{\text{min}}\ \ &c_1\sum_{n=1}^{N}\sum_{\tilde{k}_n\in\mathcal{C}_n}\left(\zeta\bar{h}_{k_n}\sqrt{\beta_{k_n,2}}\bar{h}_{\tilde{k}_n,k_n}\sqrt{\alpha_{\tilde{k}_n}}-1\right)^2\nonumber\\
			&+c_2\sum_{n=1}^{N}\sum_{ k_n\in\mathcal{C}_n}\zeta^2\nu^2\bar{h}_{k_n}^2\beta_{k_n}\sigma_{k_n}^2,\label{opt_2ndbeta:obj}\\
			{\text{s. t.}}\ \ &\beta_{k_n}\leq\hat{P}_{k_n}^{\rm max},k_n\in\mathcal{C}_n,\forall n,\label{opt_2ndbeta:bn_constraint}
		\end{align}
	\end{subequations} 
	with
	\begin{equation}
		\hat{P}_{k_n}^{\rm max}=\frac{P_{k_n}^{\rm max}}{\left(\sum_{\tilde{k}_n\in\mathcal{C}_n}\bar{h}_{\tilde{k}_n,k_n}^2\alpha_{\tilde{k}_n}+\sigma_{k_n}^2\right)}.
	\end{equation}
	Similarly, by applying the change of variables $\bar{\beta}_{k_n}\triangleq\sqrt{\beta_{k_n}}$, problem \eqref{opt_2ndbeta:problem} is equivalent to
	\begin{subequations}\label{opt_2ndbeta1:problem}
		\begin{align} 			   
			\underset{\{\bar{\beta}_{k_n}\}}{\text{min}}\ \ &c_1\sum_{n=1}^{N}\sum_{\tilde{k}_n\in\mathcal{C}_n}\left(\zeta\bar{h}_{k_n}\bar{\beta}_{k_n}\bar{h}_{\tilde{k}_n,k_n}\sqrt{\alpha_{\tilde{k}_n}}-1\right)^2\nonumber\\
			&+c_2\sum_{n=1}^{N}\sum_{k_n\in\mathcal{C}_n}\zeta^2\nu^2\bar{h}_{k_n}^2\bar{\beta}_{k_n}^2\sigma_{k_n}^2,\label{opt_2ndbeta1:obj}\\
			{\text{s. t.}}\ \ &\bar{\beta}_{k_n}\leq\sqrt{\hat{P}_{k_n}^{\rm max}},\ k_n\in\mathcal{C}_n,\forall n.\label{opt_2ndbeta1:bn_constraint}
		\end{align}
	\end{subequations}
	Obviously, problem \eqref{opt_2ndbeta1:problem} is a convex problem, and thus we obtain the optimal solution to problem \eqref{opt_2ndbeta:problem} in the following proposition.
	\begin{proposition}\label{prop:opt_beta2}
		The optimal solution to problem \eqref{opt_2ndbeta:problem}  is given as
		\begin{equation}\label{eq:optimal_beta2}
			\beta_{k_n}^{\rm opt}=\min\left[\left(\frac{c_1\zeta\bar{h}_{k_n}\sum_{\tilde{k}_n\in\mathcal{C}_n}\bar{h}_{\tilde{k}_n,k_n}\sqrt{\alpha_{\tilde{k}_n}}}{c_1\zeta^2\bar{h}_{k_n}^2\sum_{\tilde{k}_n\in\mathcal{C}_n}\bar{h}_{\tilde{k}_n,k_n}^2\alpha_{\tilde{k}_n}+c_2\zeta^2\nu^2\bar{h}_{k_n}^2\sigma_{k_n}^2}\right)^2,\hat{P}_{k_n}^{\rm max}\right], k_n\in\mathcal{C}_n,\forall n.
		\end{equation}
	\end{proposition}
	\begin{IEEEproof}
		First, we derive the optimal solution to problem \eqref{opt_2ndbeta1:problem}. By setting the first-order derivative of the objective function \eqref{opt_2ndbeta1:obj} equal to 0, we obtain
		\begin{small}
			\begin{equation}
				\begin{aligned}
					&\bar{\beta}_{k_n}^*\\
					=&\frac{c_1\zeta\bar{h}_{k_n}\sum_{\tilde{k}_n\in\mathcal{C}_n}\bar{h}_{\tilde{k}_n,k_n}\sqrt{\alpha_{\tilde{k}_n}}}{c_1\zeta^2\bar{h}_{k_n}^2\sum_{\tilde{k}_n\in\mathcal{C}_n}\bar{h}_{\tilde{k}_n,k_n}^2\alpha_{\tilde{k}_n}+c_2\zeta^2\nu^2\bar{h}_{k_n}^2\sigma_{k_n}^2}.
				\end{aligned}
			\end{equation}
		\end{small}
		and thus the optimal solution to problem \eqref{opt_2ndbeta1:problem} is given as
		\begin{equation}
			\bar{\beta}_{k_n}^{\rm opt}=\min\left[\bar{\beta}_{k_n}^*,\sqrt{\hat{P}_{k_n}^{\rm max}}\right].
		\end{equation}
		Hence, the optimal solution $\beta_{k_n}^{\rm opt}$ to problem \eqref{opt_2ndbeta:problem} is given as $\beta_{k_n}^{\rm opt}=\left(\bar{\beta}_{k_n}^{\rm opt}\right)^2$, as shown in \eqref{eq:optimal_beta2}, which completes the proof.
	\end{IEEEproof}

	\item{\textbf{Optimizing $\zeta$ for given $\{\{\alpha_{\tilde{k}_n}\},\{\beta_{k_n,}\}\}$:}} Then, problem \eqref{opt:problem} is simplified as
	\begin{align}\label{opt_zeta2:problem}		   
			\underset{\zeta>0}{\text{min}}\ \ &c_1\sum_{n=1}^{N}\sum_{\tilde{k}_n\in\mathcal{C}_n}\left(\zeta\bar{h}_{k_n}\sqrt{\beta_{k_n,2}}\bar{h}_{\tilde{k}_n,k_n}\sqrt{\alpha_{\tilde{k}_n}}-1\right)^2\nonumber\\
			&+c_2\sum_{n=1}^{N}\sum_{k_n\in\mathcal{C}_n}\zeta^2\nu^2\bar{h}_{k_n}^2\beta_{k_n}\sigma_{k_n}^2+c_2\zeta^2\nu^2\sigma^2,
	\end{align} 
	which is also a convex quadratic problem. By setting the first-order derivative of the objective function in \eqref{opt_zeta2:problem} equal to 0, the optimal solution $\zeta^{\rm opt}$ to problem \eqref{opt_zeta2:problem} is given as
	\begin{equation}\label{eq:optimal_zeta2}
		\zeta^{\rm opt}=\frac{c_1\sum_{n=1}^{N}\sum_{\tilde{k}_n\in\mathcal{C}_n}\bar{h}_{k_n}\sqrt{\beta_{k_n}}\bar{h}_{\tilde{k}_n,k_n}\sqrt{\alpha_{\tilde{k}_n}}}{\sum_{n=1}^{N}\left(c_1\sum_{\tilde{k}_n\in\mathcal{C}_n}\bar{h}_{k_n}^2\beta_{k_n}\bar{h}_{\tilde{k}_n,k_n}^2\alpha_{\tilde{k}_n}+c_2\sum_{k_n\in\mathcal{C}_n}\nu^2\bar{h}_{k_n}^2\beta_{k_n}\sigma_{k_n}^2\right)+c_2\nu^2\sigma^2}.
	\end{equation}

\end{itemize}

Hence, we summarize the alternating minimization method for solving problem \eqref{opt:problem} in Algorithm \ref{algo:optimization}. As problems \eqref{opt_alpha:problem}, \eqref{opt_2ndbeta:problem}, and \eqref{eq:optimal_zeta2} are optimally solved, it is easy to show that objective values $f^{(i)}$ are non-increasing over iterations. Hence, the convergence of Algorithm \ref{algo:optimization} is guaranteed. 

\begin{algorithm}[!htp]
	\caption{Alternating Minimization Method for Solving Problem \eqref{opt:problem}}
	\label{algo:optimization}
	\begin{algorithmic}[1]  
		\Require maximum iteration number $J_{\rm max}$ and tolerance threshold $\epsilon$.
		\State Initialize $\{\{\alpha_{\tilde{k}_n}(0)\},\{\beta_{k_n}(0)\},\zeta(0)\}$.
		\State Compute $f^{(0)}$ by substituting $\{\{\alpha_{\tilde{k}_n}(0)\},\{\beta_{k_n}\},\zeta\}$ into \eqref{opt:obj}.
		\For{$i=1,\cdots,I_{\rm max}$}
		\State With given $\{\beta_{k_n}(i)\}$ and $\zeta(i)$, obtain the optimal solution to problem \eqref{opt_alpha:problem} as $\alpha_{\tilde{k}_n}^{\rm opt}$ by Proposition \ref{prop:opt_alpha}.
		\State With given $\{\alpha_{\tilde{k}_n}^{\rm opt}\}$ and $\zeta(i)$, obtain the optimal solution to problem \eqref{opt_2ndbeta:problem} as $\beta_{k_n}^{\rm opt}$ by Proposition \ref{prop:opt_beta2}.
		\State With given $\{\alpha_{\tilde{k}_n}^{\rm opt}\}$ and $\{\beta_{k_n}^{\rm opt}\}$, obtain the optimal solution to problem \eqref{opt_zeta2:problem} as $\zeta^{\rm opt}$ by \eqref{eq:optimal_zeta2}.
		\State Update $\alpha_{\tilde{k}_n}(i)=\alpha_{\tilde{k}_n}^{\rm opt}$, $\beta_{k_n}(i)=\beta_{k_n}^{\rm opt}, \forall \tilde{k}_n, k_n\in\mathcal{C}_n,\forall n$, and $\zeta(i)=\zeta^{\rm opt}$.
		\State Update $f^{(i)}$ by substituting $\{\{\alpha_{\tilde{k}_n}(i)\},\{\beta_{k_n}(i)\},\zeta(i)\}$ into \eqref{opt:obj}.
		\If {$\frac{\left|f^{(i)}-f^{(i-1)}\right|}{\left|f^{(i-1)}\right|}\leq\epsilon$}
		\State Early stop this algorithm.
		\EndIf
		\EndFor
		\Ensure $\{\{\alpha_{\tilde{k}_n}(i)\},\{\beta_{k_n}(i)\},\zeta(i)\}$.
	\end{algorithmic}
\end{algorithm}

Based on the above analysis, the proposed AirComp-based two-tier wireless FL scheme is summarized in Algorithm \ref{algo:wireless_FL}. 
\begin{algorithm}[!htp]
	\caption{Proposed AirComp-based two-tier wireless FL Algorithm}
	\label{algo:wireless_FL}
	\begin{algorithmic}[1]  
		\Require initial global model parameter $\mathbf{w}^{(1)}$; batch size $m_b$; learning rate $\gamma$; individual transmit power budget $P_k^{\rm max}$; iteration budget $T$; and the devices' locations $\mathcal{P}$.
		\For{$t=1:T$} 
		\State PS broadcasts $\mathbf{w}^{(t)}$ to all the devices;
		\For{$k=1:K$}
		\State device $k$ estimates the local gradients by \eqref{eq:gradient} and computes the gradient statistics $\{\bar{g}_k,\nu_k^2\}$ by \eqref{eq:mean} and \eqref{eq:variance};
		\State device $k$ computes the data importance $I_k^{(t)}$ by \eqref{eq:entropy};
		\State device $k$ uploads the quantities $\{\bar{g}_k,\nu_k^2\}$ and $I_k^{(t)}$ to the PS.
		\EndFor
		\State The PS forms clusters based on device distance and data importance.
		\State The PS obtains the optimal transmit power and de-noising factor control based on Algorithm \ref{algo:optimization};
		\State The PS broadcasts the cluster result and the transmit power values to the devices;
		\For{$n=1:N$}
		\State The subordinate device $\tilde{k}_n\in\mathcal{C}_n^{(t)}$ transmits its local gradients to $k_n\in\mathcal{C}_n^{(t)}$.
		\State The lead device $k_n\in\mathcal{C}_n^{(t)}$ transmits the cluster gradients to the PS.
		\EndFor
		\State The PS aggregates the received gradients and updates the global model as $\mathbf{w}^{(t+1)}$ based on \eqref{eq:global_model};
		\EndFor
	\end{algorithmic}
\end{algorithm}

\section{Numerical Results}\label{sec:numerical}
\subsection{Simulation setup}
Here, we model all the channel coefficients as distance-dependent Rayleigh fading channels \cite{AGoldsmith,FWang} as $h_{\tilde{k}_n,k_n}^{(t)}=\sqrt{\Omega_0d_{\tilde{k}_n,k_n}^{-\kappa}}h_0$ and $h_{k_n}^{(t)}=\sqrt{\Omega_0d_{\tilde{k}_n,k_n}^{-\kappa}}h_0$, where $d_{\tilde{k}_n,k_n}$ and $d_{k_n}$ denote the distances between subordinate device $\tilde{k}_n$ and its associated lead device $k_n$ and between the lead device $k_n$ to the PS, $\Omega_0=-37$ dB is the path-loss at a reference distance of one meter, $\kappa=3.5$ denotes the path-loss exponent, and $h_0$ captures the small-scale fading of the channels and is i.i.d. across different iterations following the Rayleigh fading, i.e., $h_0\sim\mathcal{CN}(0,1)$. We consider the devices are i.i.d. distributed over a ring centered at the PS, where the inner radius is 150 m and the outer radius is 200 m\footnote{We do not consider the case that devices are uniformly distributed in the circle centered at the PS, since the devices close to the PS, i.e., the devices distributed in the nearby circle around the PS, are supposed to directly communicate to the PS without communicating to lead devices first. Hence, to focus on the considered two-tier scheme, we only consider the scenario that the devices are distributed at the network edge.}. The learning rate is set as $\gamma =\frac{1}{100L}$, where $L$ is empirically set as $L=10$ \cite{MMAmiri_1}. We consider balanced data size setting as $D_k=\frac{60000}{K}$ with $K=50$ devices. Unless stated otherwise, the number of clusters is set as $N=5$, the noise power is set as $\sigma_{k_n}^2=\sigma^2=-80$ dBm and the maximum power budget is set as $P_k^{\rm max}=0.2$ W, $\forall k$.

To achieve the image classification task, we aim to train a CNN as the classifier model, which consists of three $3\times3$ convolution layers (each with 8, 16, and 32 channels, respectively), each followed by a $2\times 2$ max pooling layer with stride 2; fully connected layer with 10 units; and finally a softmax output layer. All convolutional layers are also followed by batch normalization layers and mapped by ReLU activation. 

For performance comparison, we consider the following baseline schemes:
\begin{itemize}
	\item{Static clustering scheme:} In this scheme, the devices are divided into clusters only based on their locations. Hence, the clustering result is static across all iterations, and this clustering scheme only accounts for the communication aspect of the considered two-tier wireless FL scheme. Given the clustering result, we obtain the transmit power control and de-noising factor by Algorithm \ref{algo:optimization}.
	\item{Gradient similarity-based clustering scheme:} In this scheme, the devices are divided into clusters based on the cosine similarity of their gradients \cite{FSattler}. The clustering results are still dynamic across the training iterations, and this clustering scheme only accounts for the learning aspect of the considered two-tier wireless FL scheme. Similarly, given the clustering result, we obtain the transmit power control and de-noising factor by Algorithm \ref{algo:optimization}.
	\item{Maximum power transmissions scheme:} In this scheme, the devices use their full transmit power to transmit their gradients. For subordinate device $\tilde{k}_n\in\mathcal{C}_n,\forall n$, its transmit power is set as $\alpha_{\tilde{k}_n}=P_{\tilde{k}_n}^{\rm max}$, and the transmit power for lead device $k_n\in\mathcal{C}_n,\forall n$, is set as $\beta_{k_n}=P_{k_n}^{\rm max}/\left(\sum_{\tilde{k}_n\in\mathcal{C}_n}\bar{h}_{\tilde{k}_n,k_n}^2\alpha_{\tilde{k}_n}+\sigma_{k_n}^2\right)$.
	\item{Direct gradient transmission scheme:} In this scheme, the devices directly transmit their gradients to the PS based on AirComp, and the transmit power and de-noising factor are obtained by solving the corresponding optimization problem similar to Problem \eqref{opt:problem}.
	\item{Conventional MSE minimization scheme:} In this scheme, we directly minimize the MSE w.r.t. the normalized symbol $\mathbf{s}_k^{(t)}$ \cite{ZLin,FWang}. 
\end{itemize}

\subsection{I.i.d. data distribution case}
In this subsection, we evaluate the performance of the proposed scheme for the i.i.d. data distribution case, where we randomly assign the training samples to each device.

\begin{figure}[H] 
	\centering 
	\includegraphics[width=0.5\textwidth]{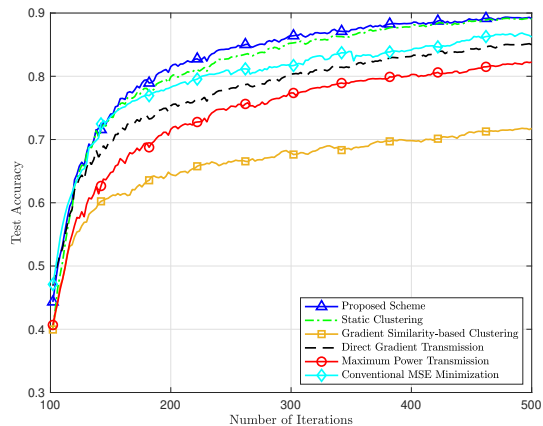} 
	\caption{Test accuracy vs. number of iterations with i.i.d. data distribution.}
	\label{fig:conv_iid}
\end{figure}

Fig. \ref{fig:conv_iid} compares the performances of all the schemes w.r.t. the iteration in terms of test accuracy with i.i.d. data distribution. It is observed that all the schemes converge after sufficiently larger number of iterations. When converges, the test accuracy of the proposed scheme is higher than those of the rest baseline schemes, while it performs close to the static clustering scheme. The reason why the proposed and static clustering schemes achieve close learning performance is that the data importance of the devices are quite close with i.i.d. data distribution. Hence, devices' locations dominate the clustering objective when applying the hierarchical clustering method and the proposed clustering method is nearly static during the training process.

\begin{figure}[H] 
	\centering 
	\includegraphics[width=0.5\textwidth]{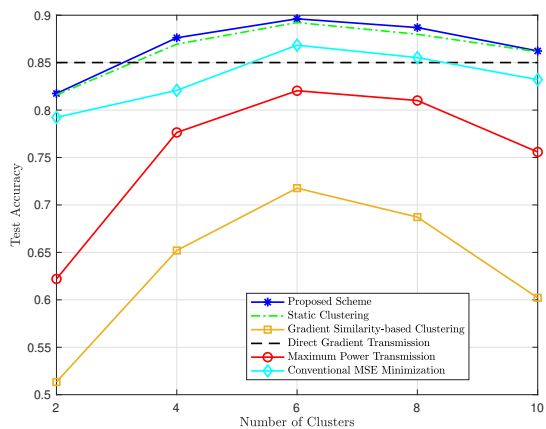} 
	\caption{Test accuracy vs. number of clusters with i.i.d. data distribution.}
	\label{fig:cluster_iid}
\end{figure}

Fig. \ref{fig:cluster_iid} compares the performances of all the schemes w.r.t. the number of cluster (i.e., $N$) in terms of the average test accuracy when they converge with i.i.d. data distribution. It is observed that all the accuracies of the schemes that adopt the two-tier structures (i.e., excluding the direct gradient transmission scheme) first increase and then decrease with the number of clusters. The reason why these schemes suffer from performance losses when $N$ is small is that lead devices are required to aggregate a large number of subordinate devices, even when the subordinate devices are far away from the associated lead devices. In this case, the subordinate devices with relatively large distances to the lead device suffer from bad channel fading, and the transmit power of the lead device is limited since they need to aggregate too many gradients from the subordinate devices. The reason why these schemes suffer from performance losses when the number of cluster becomes large is that larger $N$ results in less data exploitation, since the lead devices only account for relaying the aggregated gradients and do not upload their gradients to the PS. For different numbers of clusters, i.e., $2\leq N\leq 10$, the static clustering scheme achieves close learning performance to the proposed scheme due the same reason as explained in Fig. \ref{fig:conv_iid}. The accuracy of the proposed scheme is always much higher than the gradient-similarity based clustering, maximum transmission power, and conventional MSE schemes, and is higher than the direct gradient transmission scheme when $N>2$.

\begin{figure}[H] 
	\centering 
	\includegraphics[width=0.5\textwidth]{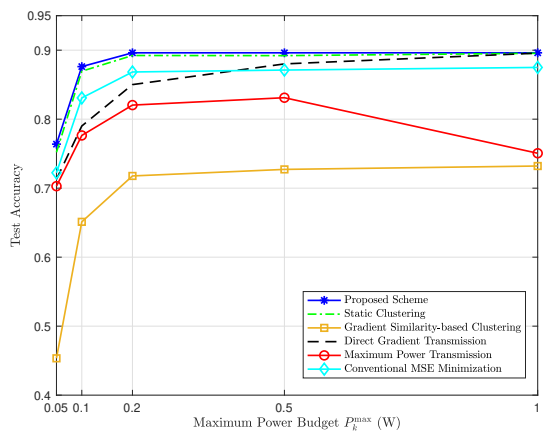} 
	\caption{Test accuracy vs. maximum transmit power budget with i.i.d. data distribution.}
	\label{fig:budget_iid}
\end{figure}

Fig. \ref{fig:budget_iid} compares the performances of all the schemes w.r.t. different levels of maximum transmit power budget $P_k^{\rm max}$, in terms of the average test accuracy when they converge with i.i.d. data distribution. It is observed that the test accuracies of all schemes excluding the maximum power transmit scheme increase with with the maximum transmit power budget, while the test accuracy of the maximum power transmission scheme first increases and then decreases with the maximum transmit power budget. This is due to fact that gradient misalignment error is quite large when the devices always transmit with sufficiently large transmit power. Obviously, the proposed scheme outperforms the rest baseline schemes and the static clustering scheme achieves close performance to the proposed scheme for different levels of $P_k^{\rm max}$.  Besides, with sufficiently large transmit power budget, i.e., $P_k^{\rm max}=1$ W, the direct gradient transmission scheme achieves comparable test accuracy to the proposed scheme. This is because when $P_k^{\rm max}$ is sufficiently large, both these two schemes exhibit optimal channel inversion power control and the noise-induced errors can be suppressed into relatively low levels for both of them due to the de-noising factor. 

\subsection{Non-i.i.d. data distribution case}
In this subsection, we evaluate the performance of the proposed scheme for the non-i.i.d. data distribution case,  where we split the training samples into 5 disjoint subsets with each subset containing 2 classes of images, and each subset is chose to randomly assign training samples to $K/5$ devices. Hence, each device only contains two classes of images and different devices contains different classes of images.

\begin{figure}[H] 
	\centering 
	\includegraphics[width=0.5\textwidth]{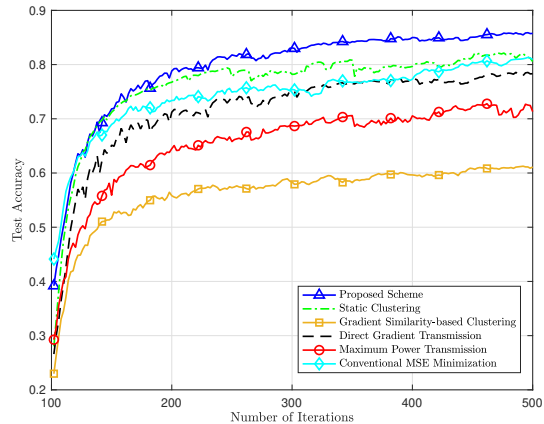} 
	\caption{Test accuracy vs. number of iterations with non-i.i.d. data distribution.}
	\label{fig:conv_noniid}
\end{figure}

\begin{figure}[H] 
	\centering 
	\includegraphics[width=0.5\textwidth]{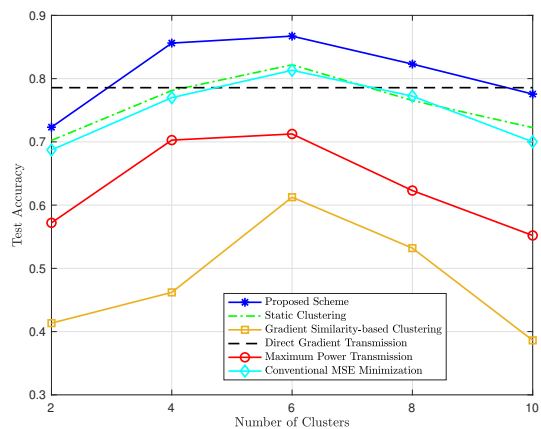} 
	\caption{Test accuracy vs. number of clusters with non-i.i.d. data distribution.}
	\label{fig:cluster_noniid}
\end{figure}

Fig. \ref{fig:conv_noniid} compares the performances of all the schemes w.r.t. the number of iterations in terms of test accuracy with non-i.i.d. data distribution. It is observed that all the schemes still converge after sufficient large number of iterations and the proposed scheme also outperforms the baseline schemes in terms of convergence rate and test accuracy. Besides, compared to the i.i.d data distribution case in \eqref{fig:conv_iid}, all the schemes suffer from performance loss, and the test accuracy for the proposed scheme when converges is significantly higher than that of the static clustering scheme, which shows the effectiveness of the joint consideration of device distance and data importance for clustering under the non-i.i.d. data distribution case. Besides, the increasing trends of all the schemes become less stable compared to those in the i.i.d. data distribution case, which is obviously caused by the non-i.i.d. data distribution.

\begin{figure}[H] 
	\centering 
	\includegraphics[width=0.5\textwidth]{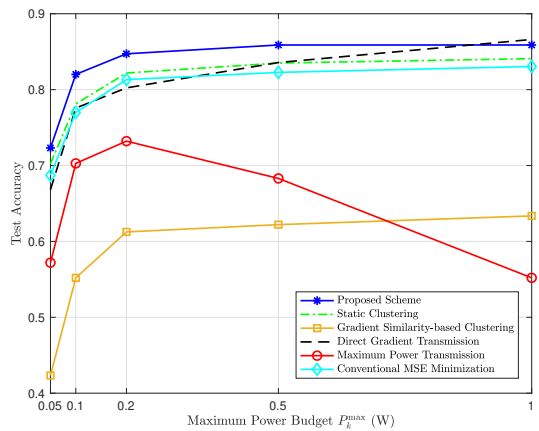} 
	\caption{Test accuracy vs. maximum transmit power budget with non-i.i.d. data distribution.}
	\label{fig:budget_noniid}
\end{figure}

Fig. \ref{fig:cluster_noniid} compares the performances of all the schemes w.r.t. different number of cluster $N$ in terms of the average test accuracy when they converge with non-i.i.d. data distribution. Similar to the i.i.d. data distribution case, the accuracies of all the schemes that adopt the two-tier structures first increase and then decrease with the number of clusters and the accuracy of the proposed scheme is always much higher than the rest two-tier baseline schemes. When the number of clusters is quite small, i.e., $N=2$, the accuracy of the proposed scheme is slightly lower than that of the direct gradient transmission scheme due to the same reason as explained for the case shown in Fig. \ref{fig:cluster_iid}. When the number of clusters $N$ becomes large, i.e., $N=10$, the proposed scheme not only suffer from performance loss compared to the case when $N$ is moderate, i.e., $4\leq N\leq8$, but also has lower test accuracy than the direct gradient transmission scheme, which is different to the case with i.i.d. data distribution case in Fig. \ref{fig:cluster_iid}. This is due to the fact that the negative effect of insufficient data exploitation becomes much more severe for the non-i.i.d. data distribution case. 

Fig. \ref{fig:budget_noniid} compares the performances of all the schemes w.r.t. different levels of maximum transmit power budget $P_k^{\rm max}$ in terms of the average test accuracy, when they converge with non-i.i.d. data distribution. Similar to the i.i.d. data distribution case in Fig. \ref{fig:budget_iid}, the test accuracies of all the proposed schemes excluding the maximum power transmission scheme increase with the maximum transmit power budget. The test accuracy of the maximum power transmission scheme first increases and then decreases with the maximum power budget, and the test accuracy with sufficiently large power budget is even lower than that with small power budget. This is due to the negative effects of both the large gradient misalignment error with sufficiently high transmit power and insufficient data exploitation with non-i.i.d data distribution. Besides, different from the case with i.i.d. data distribution, with sufficiently large transmit power budget, i.e., $P_k^{\rm max}=1$ W, the direct gradient transmission scheme has slightly higher test accuracy compared to the proposed scheme. This is because the proposed scheme still suffers from negative effect of insufficient data exploitation while the direct gradient transmission scheme does not. Such negative effect is much more severe in the non-i.i.d. data distribution case compared to the i.i.d. data distribution case. However, the proposed scheme outperforms all the baseline schemes when the transmit power budget is not sufficiently large enough.

\section{Conclusion}\label{sec:conclusion}
In this paper, we studied the joint communication and learning design for an AirComp-based two-tier wireless FL scheme, where all devices are dynamically partitioned into different clusters across different training iterations and the PS aggregates the gradients from the devices through AirComp-based intra-cluster and inter-cluster gradient aggregations. First, we analyzed the convergence behavior of the proposed scheme and the optimality gap between the expected and optimal global loss values was derived under mild conditions. The optimality gap characterized the impact of the AirComp-based two-tier gradient aggregation errors on the convergence of the proposed scheme, which needs to be minimized to enhance the performance of the proposed scheme. To this end, we proposed the clustering method and studied optimality gap minimization problem. By jointly considering the device distance and data importance as the minimax linkage criterion, we proposed a low-complexity hierarchical clustering method that can be efficiently implemented during the training process. Then, with the instantaneous CSI and clustering result in each iteration, we formulated the optimality gap minimization problem under the individual transmit power constraint. To solve the optimization problem, we proposed an alternating minimization method, where the optimal solution for the subordinate devices follows regularized composite channel inversion structure. Finally, numerical results showed that our proposed scheme outperforms the baseline schemes. 
\appendices
\section{Proof of Theorem \ref{th:opt_gap}}\label{appen:th1_proof}
First, we introduce the following lemma.
\begin{lemma}\label{lem:bounded_global_gradient}
The expectation of the square norm of the gradient of the global loss function at each round is bounded by
\begin{equation}\label{eq:lem}
	\|\nabla F(\mathbf{w}^{(t)})\|_2^2\leq 2L\left(F(\mathbf{w}^{(t)})-F^*\right),
\end{equation}
where $F^*$ is the optimal global loss function value.
\end{lemma}
\begin{IEEEproof}
	From \eqref{eq:smooth_1}, we obtain
	\begin{align}\label{proof:eq13}
		F(\mathbf{w}(t))-F(\mathbf{w}^*)\leq&\nabla F(\mathbf{w}^*)(\mathbf{w}(t)-\mathbf{w}^*)+\frac{L}{2}\|\mathbf{w}(t)-\mathbf{w}^*\|_2^2,\nonumber\\
		&\overset{(a)}{=}\frac{L}{2}\|\mathbf{w}(t)-\mathbf{w}^*\|_2^2,
	\end{align}	
	where $(a)$ is due to $\nabla F(\mathbf{w}^*)=0$, since $\mathbf{w}^*$ is the optimal solution. Besides, we have
	\begin{align}\label{proof:eq14}
		F(\mathbf{w}^*)=&\min_{\mathbf{w}} F(\mathbf{w})\nonumber\\
		\leq&\min_{\mathbf{w}}\left[F(\mathbf{w}(t))+\nabla F(\mathbf{w}(t))^T(\mathbf{w}-\mathbf{w}(t))+\frac{L}{2}\|\mathbf{w}-\mathbf{w}(t)\|_2^2\right]\nonumber\\
		=&F(\mathbf{w}(t))+\min_{\|\mathbf{a}\|_2=1}\min_{h\geq 0}\left[F(\mathbf{w}(t))^T\mathbf{a}h+\frac{L}{2}h^2\right]\nonumber\\
		=&F(\mathbf{w}(t))+\min_{\|\mathbf{a}\|_2=1}\left[-\frac{(F(\mathbf{w}(t))^T\mathbf{a})^2}{2L}\right]\nonumber\\
		=&F(\mathbf{w}(t))-\frac{\|F(\mathbf{w}(t))\|_2^2}{2L},
	\end{align}
	which yields \eqref{eq:lem} in Lemma \ref{lem:bounded_global_gradient}. This thus completes the proof.
\end{IEEEproof}
Then, according to \eqref{eq:smooth_1} and \eqref{eq:new_glob_model}, we have
\begin{align}\label{eq:proof_eq1}
	F(\mathbf{w}^{(t+1)})\leq& F(\mathbf{w}^{(t)})+\langle\mathbf{w}^{(t+1)}-\mathbf{w}^{(t)},\nabla F\left(\mathbf{w}^{(t)}\right)\rangle+\frac{L}{2}\|\mathbf{w}^{(t+1)}-\mathbf{w}^{(t)}\|_2^2\nonumber\\
	=&F(\mathbf{w}^{(t)})-\left\langle\mathbf{g}^{(t)}+\bm{\varepsilon}^{(t)},\nabla F(\mathbf{w}^{(t)})\right\rangle+\frac{L\gamma^2}{2}\left\|\mathbf{g}^{(t)}+\bm{\varepsilon}^{(t)}\right\|_2^2,
\end{align}
and thus it follows 
\begin{align}\label{eq:proof_eq2}
	F(\mathbf{w}^{(t+1)})-F(\mathbf{w}^{(t)})\leq-\gamma\left\langle\mathbf{g}^{(t)}+\bm{\varepsilon}^{(t)},\nabla F(\mathbf{w}^{(t)})\right\rangle+\frac{L\gamma^2}{2}\left\|\mathbf{g}^{(t)}+\bm{\varepsilon}^{(t)}\right\|_2^2,
\end{align}
By taking the expectation on both sides of \eqref{eq:proof_eq2}, we obtain
\begin{align}\label{eq:proof_eq3}
	&\mathbb{E}\left[F(\mathbf{w}^{(t+1)})-F(\mathbf{w}^{(t)})\right]\nonumber\\
	\leq&-\gamma\left\langle\mathbb{E}\left[\mathbf{g}^{(t)}+\bm{\varepsilon}^{(t)}\right],\nabla F(\mathbf{w}^{(t)})\right\rangle+\frac{L\gamma^2}{2}\mathbb{E}\left[\left\|\mathbf{g}^{(t)}+\bm{\varepsilon}^{(t)}\right\|_2^2\right]\nonumber\\
	=&-\gamma\left\langle\mathbb{E}\left[\mathbf{g}^{(t)}\right],\nabla F(\mathbf{w}^{(t)})\right\rangle-\gamma\left\langle\mathbb{E}\left[\bm{\varepsilon}^{(t)}\right],\nabla F(\mathbf{w}^{(t)})\right\rangle+\frac{L\gamma^2}{2}\mathbb{E}\left[\left\|\mathbf{g}^{(t)}+\bm{\varepsilon}^{(t)}\right\|_2^2\right]\nonumber\\
	\overset{(a)}{=}&-\gamma\|\nabla F(\mathbf{w}^{(t)})\|^2-\gamma\left\langle\mathbb{E}\left[\bm{\varepsilon}^{(t)}\right],\nabla F(\mathbf{w}^{(t)})\right\rangle+\frac{L\gamma^2}{2}\mathbb{E}\left[\left\|\mathbf{g}^{(t)}+\bm{\varepsilon}^{(t)}\right\|_2^2\right]\nonumber\\
	\overset{(b)}{\leq}&-\gamma\|\nabla F(\mathbf{w}^{(t)})\|^2-\gamma\left\langle\mathbb{E}\left[\bm{\varepsilon}^{(t)}\right],\nabla F(\mathbf{w}^{(t)})\right\rangle+L\gamma^2\mathbb{E}\left[\left\|\mathbf{g}^{(t)}\right\|_2^2\right]+L\gamma^2\mathbb{E}\left[\|\bm{\varepsilon}^{(t)}\|^2\right]\nonumber\\
	\overset{(c)}{\leq}&-\frac{\gamma}{2}\|\nabla F(\mathbf{w}^{(t)})\|^2+\frac{\gamma}{2}\left\|\mathbb{E}\left[\bm{\varepsilon}^{(t)}\right]\right\|^2+L\gamma^2\mathbb{E}\left[\left\|\mathbf{g}^{(t)}\right\|_2^2\right]+L\gamma^2\mathbb{E}\left[\|\bm{\varepsilon}^{(t)}\|^2\right]\nonumber\\
	\overset{(d)}{\leq}&-\frac{\gamma}{2}\|\nabla F(\mathbf{w}^{(t)})\|^2+\frac{\gamma}{2}\left\|\mathbb{E}\left[\bm{\varepsilon}^{(t)}\right]\right\|^2+\frac{L\gamma^2}{K}\sum_{k=1}^{K}\mathbb{E}\left[\left\|\mathbf{g}_k(t)\right\|_2^2\right]+L\gamma^2\mathbb{E}\left[\|\bm{\varepsilon}^{(t)}\|^2\right]\nonumber\\
	\overset{(e)}{\leq}&-\frac{\gamma}{2}\|\nabla F(\mathbf{w}^{(t)})\|^2+\frac{\gamma}{2}\left\|\mathbb{E}\left[\bm{\varepsilon}^{(t)}\right]\right\|^2+L\gamma^2\left\|\nabla F(\mathbf{w}^{(t)})\right\|_2^2+\frac{L\gamma^2}{m_b}\|\bm{\sigma}\|^2+L\gamma^2\mathbb{E}\left[\|\bm{\varepsilon}^{(t)}\|^2\right]\nonumber\\
	=&\frac{2L\gamma^2-\gamma}{2}\|\nabla F(\mathbf{w}^{(t)})\|^2+\frac{L\gamma^2}{m_b}\|\bm{\delta}\|^2+\frac{\gamma}{2}\left\|\mathbb{E}\left[\bm{\varepsilon}^{(t)}\right]\right\|^2+L\gamma^2\mathbb{E}\left[\|\bm{\varepsilon}^{(t)}\|^2\right],
\end{align}
where $(a)$ follows \eqref{eq:unbiasedSGD} in Assumption \ref{as:bounded_sgd}, $(b)$ follows the inequality $\|\mathbf{a}_1+\mathbf{a}_2\|_2^2\leq2\|\mathbf{a}_1\|_2^2+2\|\mathbf{a}_2\|_2^2$, $(c)$ follows the inequality of arithmetic and geometric means, i.e., $-\langle\mathbf{a}_1,\mathbf{a}_2\rangle\leq\frac{\|\mathbf{a}_1\|_2^2}{2}+\frac{\|\mathbf{a}_2\|_2^2}{2}$, $(d)$ follows the Jensen's inequality, and $(e)$ follows \eqref{eq:boundedSGD} in Assumption \ref{as:bounded_sgd}.

Next, substituting \eqref{eq:lem} into the RHS of \eqref{eq:proof_eq3}, we obtain
\begin{align}\label{eq:proof2_eq1}
	&\mathbb{E}\left[F(\mathbf{w}^{(t+1)})-F(\mathbf{w}^{(t)})\right]\nonumber\\
	\leq&\frac{2L\gamma^2-\gamma}{2}\|\nabla F(\mathbf{w}^{(t)})\|^2+\frac{L\gamma^2}{m_b}\|\bm{\sigma}\|^2+\frac{\gamma}{2}\left\|\mathbb{E}\left[\bm{\varepsilon}^{(t)}\right]\right\|^2+L\gamma^2\mathbb{E}\left[\|\bm{\varepsilon}^{(t)}\|^2\right]\nonumber\\
	\leq&(2L^2\gamma^2-L\gamma)\left(F(\mathbf{w}^{(t)})-F^*\right)+\frac{L\gamma^2}{m_b}\|\bm{\sigma}\|^2+\frac{\gamma}{2}\left\|\mathbb{E}\left[\bm{\varepsilon}^{(t)}\right]\right\|^2+L\gamma^2\mathbb{E}\left[\|\bm{\varepsilon}^{(t)}\|^2\right].
\end{align}
Through some further algebraic manipulation, we obtain
\begin{align}\label{eq:proof2_eq2}
	\mathbb{E}\left[F(\mathbf{w}^{(t+1)})\right]-F^*\leq&(2L^2\gamma^2-L\gamma+1)\left(\mathbb{E}\left[F(\mathbf{w}^{(t)})\right]-F^*\right)\nonumber\\
	&+\frac{L\gamma^2}{m_b}\|\bm{\sigma}\|^2+\frac{\gamma}{2}\left\|\mathbb{E}\left[\bm{\varepsilon}^{(t)}\right]\right\|^2+L\gamma^2\mathbb{E}\left[\|\bm{\varepsilon}^{(t)}\|^2\right].
\end{align}
Assume the FL algorithm terminates after $T$ rounds. Given an initial global model $\mathbf{w}^{(1)}$, by carrying out recursions for \eqref{eq:proof2_eq2}, Theorem \ref{th:opt_gap} is proved.

\section{Proof of Proposition \ref{prop:opt_alpha}}\label{appen:prop_alpha}
Since problem \eqref{opt_alpha1:problem} is convex and satisfies the Slater's condition, strong duality holds between problem \eqref{opt_alpha1:problem} and its Lagrange dual problem \cite{SBoyd}. Let $\lambda_{\tilde{k}_n}$ and $\mu_{n}$ denote the dual variables associated with the $\tilde{k}_n$-th constraint in \eqref{opt_alpha1:ak_constraint} and the $n$-th constraint in \eqref{opt_alpha1:bn_constraint}, respectively. Hence, the Lagrange of of problem \eqref{opt_alpha1:problem} is given as
\begin{small}
	\begin{align}\label{eq:lagrange}
		\mathcal{L}
		=&\sum_{n=1}^{N}\sum_{\tilde{k}_n\in\mathcal{C}_n}\left(c_1\left(\zeta\bar{h}_{k_n}\sqrt{\beta_{k_n}}\bar{h}_{\tilde{k}_n,k_n}\bar{\alpha}_{\tilde{k}_n}-1\right)^2+\lambda_{\tilde{k}_n}\bar{\alpha}_{\tilde{k}_n}\right)+\sum_{n=1}^{N}\sum_{\tilde{k}_n\in\mathcal{C}_n,\tilde{k}_n\neq k_n}\mu_{n}\bar{h}_{\tilde{k}_n,k_n}^2\bar{\alpha}_{\tilde{k}_n}^2+\sum_{n=1}^{N}\mu_{n}\sigma_{k_n}^2\nonumber\\
		&-\sum_{n=1}^{N}\sum_{\tilde{k}_n\in\mathcal{C}_n}\lambda_{\tilde{k}_n}\sqrt{P_{\tilde{k}_n}^{\rm max}}-\sum_{n=1}^{N}\mu_{n}\bar{P}_{k_n}^{\rm max}.
	\end{align}
\end{small}
Then, the optimal primal and dual variables satisfy the Karush-Kuhn-Tucker (KKT) optimality conditions, which are given as
\begin{align}\label{eq:kkt}
	&\bar{\alpha}_{\tilde{k}_n}^*\leq\sqrt{P_{\tilde{k}_n^{\rm max}}},\ \tilde{k}_n\in\mathcal{C}_n,\forall n,\\
	&\sum_{\tilde{k}_n\in\mathcal{C}_n}\bar{h}_{\tilde{k}_n,k_n}^2\left(\bar{\alpha}_{\tilde{k}_n}^*\right)^2+\sigma_{k_n}^2\leq \bar{P}_{k_n}^{\rm max},\forall n,\\
	&\lambda_{\tilde{k}_n}^*\geq0,\forall \tilde{k}_n\in\mathcal{K}\setminus\mathcal{K}_N,\mu_{n}^*\geq 0,n=1,\cdots,N,\\
	&\lambda_{\tilde{k}_n}^*\left(\bar{\alpha}_{\tilde{k}_n}^*-\sqrt{P_{\tilde{k}_n^{\rm max}}}\right)=0, \tilde{k}_n\in\mathcal{C}_n,\forall n,\\
	&\mu_{n}^*\left(\sum_{\tilde{k}_n\in\mathcal{C}_n}\bar{h}_{\tilde{k}_n,k_n}^2\left(\bar{\alpha}_{\tilde{k}_n}^*\right)^2+\sigma_{k_n}^2-\bar{P}_{k_n}^{\rm max}\right)=0,\forall n,\label{eq:mu_constraint}\\
	&\frac{\partial \mathcal{L}}{\partial\bar{\alpha}_{\tilde{k}_n}^*}=\left(2c_1\zeta^2\bar{h}_{k_n}^2\beta_{k_n}\bar{h}_{\tilde{k}_n,k_n}^2+2\mu_{n}^*\bar{h}_{\tilde{k}_n,k_n}^2\right)\bar{\alpha}_{\tilde{k}_n}^*+\lambda_{\tilde{k}_n}^*-2c_1\zeta\bar{h}_{k_n}\sqrt{\beta_{k_n}}\bar{h}_{\tilde{k}_n,k_n}=0,\tilde{k}_n\in\mathcal{C}_n,\forall n.
\end{align}
Based on the KKT optimality conditions, we obtain the optimal solution $\bar{\alpha}_{\tilde{k}_n}^*$ to problem \eqref{opt_alpha1:problem} as 
\begin{align}
	\bar{\alpha}_{\tilde{k}_n}^*=\min\left[\frac{c_1\zeta\bar{h}_{k_n}\sqrt{\beta_{k_n,2}}\bar{h}_{\tilde{k}_n,k_n}}{c_1\zeta^2\bar{h}_{k_n}^2\beta_{k_n}\bar{h}_{\tilde{k}_n,k_n}^2+\mu_{n}^*\bar{h}_{\tilde{k}_n,k_n}^2}, \sqrt{P_{\tilde{k}_n}^{\rm max}}\right],
\end{align}
where non-negative $\{\mu_{n}^*\}$ satisfies the complementary slackness conditions in \eqref{eq:mu_constraint}. Hence, the optimal solution $\alpha_{\tilde{k}_n}^{\rm opt}$ to problem \eqref{opt_alpha:problem} is given as $\alpha_{\tilde{k}_n}^{\rm opt}=\left(\bar{\alpha}_{\tilde{k}_n}^*\right)^2$, as shown in \eqref{eq:optimal_alpha}. This thus completes the proof.

\end{document}